\documentclass[%
reprint,
superscriptaddress,
floatfix,
amsmath,amssymb,
aps,
prl,
]{revtex4-2}

\usepackage{graphicx}
\usepackage{dcolumn}
\usepackage{bm}

\usepackage{xcolor}

\newcommand{\beginsupplement}{%
	\setcounter{table}{0}
	\renewcommand{\thetable}{S\arabic{table}}%
	\setcounter{figure}{0}
	\renewcommand{\thefigure}{S\arabic{figure}}%
	\renewcommand{\theequation}{S.\arabic{equation}}
}

\begin{document}
	
	\preprint{APS/123-QED}
	
	\title{Sharpness of the Berezinskii-Kosterlitz-Thouless transition in disordered NbN films}
	
	\author{Alexander Weitzel}
	\altaffiliation[]{These authors contributed equally.}
	\affiliation{
		Institute for Experimental and Applied Physics, University of Regensburg, D-93040 Regensburg, Germany}
	
	\author{Lea Pfaffinger}%
	\altaffiliation[]{These authors contributed equally.}
	\affiliation{
		Institute for Experimental and Applied Physics, University of Regensburg, D-93040 Regensburg, Germany}

	\author{Ilaria Maccari}%
	\affiliation{
		Department of Physics, Stockholm University, SE-10691 Stockholm, Sweden}

	\author{Klaus Kronfeldner}
	\affiliation{
		Institute for Experimental and Applied Physics, University of Regensburg, D-93040 Regensburg, Germany}
	
	\author{Thomas Huber}
	\affiliation{
		Institute for Experimental and Applied Physics, University of Regensburg, D-93040 Regensburg, Germany}
	
	\author{Lorenz Fuchs}
	\affiliation{
		Institute for Experimental and Applied Physics, University of Regensburg, D-93040 Regensburg, Germany}
	
		\author{James Mallord}
	\affiliation{
		Institute for Experimental and Applied Physics, University of Regensburg, D-93040 Regensburg, Germany}
	
	\author{Sven Linzen}
	\affiliation{
		Leibniz Institute of Photonic Technology, D-07745 Jena, Germany}
	
	\author{Evgeni Il'ichev}
	\affiliation{
		Leibniz Institute of Photonic Technology, D-07745 Jena, Germany}
	
	\author{Nicola Paradiso}
	\affiliation{
		Institute for Experimental and Applied Physics, University of Regensburg, D-93040 Regensburg, Germany}
	
	\author{Christoph Strunk}
	\affiliation{
		Institute for Experimental and Applied Physics, University of Regensburg, D-93040 Regensburg, Germany}

	
	%
	
	\date{\today}

\begin{abstract}
		We present a comprehensive investigation of the Berezinskii-Kosterlitz-Thouless (BKT) transition in ultrathin strongly disordered NbN films. Measurements of resistance, current-voltage characteristics and kinetic inductance on the very same device reveal a consistent picture of a sharp unbinding transition of vortex-antivortex pairs that fit standard renormalization group theory without extra assumptions in terms of inhomogeneity. Our experiments demonstrate that the previously observed broadening of the transition is not an intrinsic feature of strongly disordered superconductors and provide a clean starting point for the study of dynamical effects at the BKT transition.  

\end{abstract}

\maketitle

In two dimensions, the superfluid transition is governed by the presence of thermally excited vortex-antivortex pairs \cite{KT_1973, Kosterlitz_1974}.  For superfluid $^4$He films, the defining features of the Berezinskii-Kosterlitz-Thouless (BKT) transition are well understood \cite{NelsonKosterlitz1977,McQueeney1984}. In thin-film superconductors an analogous behavior is expected, the transition being caused by dissociation of vortex-antivortex pairs. The transition is manifested as a discontinuous jump in the superfluid phase stiffness $J_s$ at a temperature $T_\mathrm{BKT}$ below the mean-field transition temperature $T_\mathrm{c0}$ \cite{AHNS1,AHNS2}. Moreover, below $T_\mathrm{BKT}$ the voltage-current characteristics $V(I)$ are nonlinear, $V\propto I^{\alpha(T)}$, with a temperature dependent exponent $\alpha$  \cite{HalperinNelson_1979}. In the thermodynamic limit, a linear voltage response regime exists above $T_\mathrm{BKT}$ only. 

Physics of the BKT-transition is controlled by two energy scales \cite{Benfatto2009}. In order to thermally excite a vortex-antivortex pair in a film, the energy cost for the generation of vortex cores (also called vortex fugacity) $\mu \propto \xi^2$ as well as the energy scale for the pair dissociation $J_s \propto 1/\lambda^2$ must be sufficiently small. Here $\xi$ and $\lambda$ are the coherence length and magnetic penetration depth, respectively.  In the dirty limit, both $\xi^2$ and $1/\lambda^2$ are proportional to the elastic mean free path. Owing to their small $\mu$ and $J_s$, ultrathin films of strongly disordered superconductors are the preferred choice for materials that feature a large separation between $T_\mathrm{BKT}$ and the mean field critical temperature $T_{c0}$.  

In the past $J_s(T)$ and $V(I)$ were studied for InO and NbN thin films \cite{FioryHebardGlaberson_1983, Yong2013, Venditti2019} using the two-coil method \cite{Turneaure2000} and standard transport measurements. The two-coil method requires circular films with typical 10\,mm diameter, while for dc-transport long strips are needed. Hence,  $J_s(T)$ and $R(T)$ could not be studied in the same devices limiting the validity of consistency checks. While a qualitative agreement with original theory was observed, measurements of strongly disordered NbN-films always displayed a broadening of the BKT-transition, far stronger than expected for, e.g., finite size effects alone \cite{Benfatto2009, Mondal_2011b}. At present, such broadening is believed to be typical for highly disordered superconducting films that are known to feature \textit{emergent granularity} \cite{Ghosal1998,Ghosal2001b, Sacepe2008, Carbillet2016, Carbillet2020, Stosiek2020}. Local variations of the modulus of the order parameter and superfluid stiffness could, in principle, explain the observed smearing of the expected discontinuous jump in $J_s$. On the other hand, such smearing introduces an additional free parameter that inevitably obscures the quantitative analysis.

Within the generally accepted picture, individual signatures of the BKT transition have been observed \cite{Mondal_2011b, Yong2013, FioryHebardGlaberson_1983, Turneaure2000, Crane2007, Mandal2020, Broun2007, Kamal1994, Yong2012, Zuev2005, Maccari2017, Venditti2019, Medveyeva2000,Ganguly2015, Mallik2022}.  In recent years, however, it turned out that each of these signatures is affected by experimental subtleties that need to be controlled in order to reliably test the level of  consistency \cite{Tamir2019, Benyamini2020}. The most popular signature, the non-linearity of $V(I)$, is also the most difficult to interpret, as many other effects affect it. For example, any fluctuation induced broadening of the resistive transition leads to non-linear $V(I)$ via heating. This can mimic a power-law behavior, in particular close to the normal state resistance and for materials with $T_\mathrm{c0}\lesssim1\,$K \cite{Levinson2019}. To address this issue, a set of techniques is desirable that do not extrinsically broaden the transition and allows for all types of measurements to be performed on the very same device.

In this Letter, we observe a sharp BKT-transition in strongly disordered NbN films while the resistive transition is smeared over several kelvins. Using a low-frequency resonator technique compatible with four terminal DC-measurements, we unambiguously identify the BKT- and mean field transition temperatures. We find an excellent agreement of both $T_\mathrm{BKT}$  and $T_{c0}$  extracted from DC-resistance and superfluid stiffness in disjunct temperature regimes.  The inductively measured stiffness shows excellent agreement with the values extracted from non-linear DC-transport, provided that voltages are sufficiently small. %
Our results provide a solid basis for the study of more complex non-equilibrium properties of ultra-thin and strongly disordered superconductors.

Our NbN films are grown by atomic layer deposition (ALD - 75 cycles) \cite{Linzen_2017} with a thickness $d = 3.5\pm0.3$~nm on top of a thermally oxidized silicon wafer. Over several months at ambient conditions, the NBN-film gradually oxidizes, signaled by an increase of the sheet resistance. Using standard electron beam lithography and selective etching techniques we prepared long ($\sim$ 100-200 squares) meander structures of width ranging from 10 - 200~\textmu m with a total kinetic inductance of $\sim100$\,nH. The samples are mounted into a cold RLC circuit, whose resonance frequency provides access to the sheet kinetic inductance $L_\square$ of the sample \cite{Baumgartner_2020,Supplement}. The resonance frequency of the circuit varies between 0.5\,-\,3\,MHz, depending on $L_\square$.
From the kinetic inductance, the superfluid stiffness is inferred as 
\begin{align}\label{eq:J_S}
	J_s\ =\ \frac{\hbar^2d}{4e^2k_\mathrm{B}\mu_0\lambda^2}\ =\ \frac{\hbar^2}{4e^2k_\mathrm{B} L_{\square}}\;, 
\end{align}
where  $h$ is the Planck's constant, $e$ the electron charge and $k_\mathrm{B}$ being Boltzmann's constant. 
The parameters of our films are well in line with those of \cite{Mondal_2011a,Mondal_2011b}, albeit with lower thickness for the same values of  $k_F\ell$ and $T_\mathrm{c0}$. Additional voltage probes allow for measurement of DC $V(I)$ characteristics on the same device. Resistance values were always extracted from the linear regime of $V(I)$. 

\begin{figure}[t]
	\includegraphics[width=.45\textwidth]{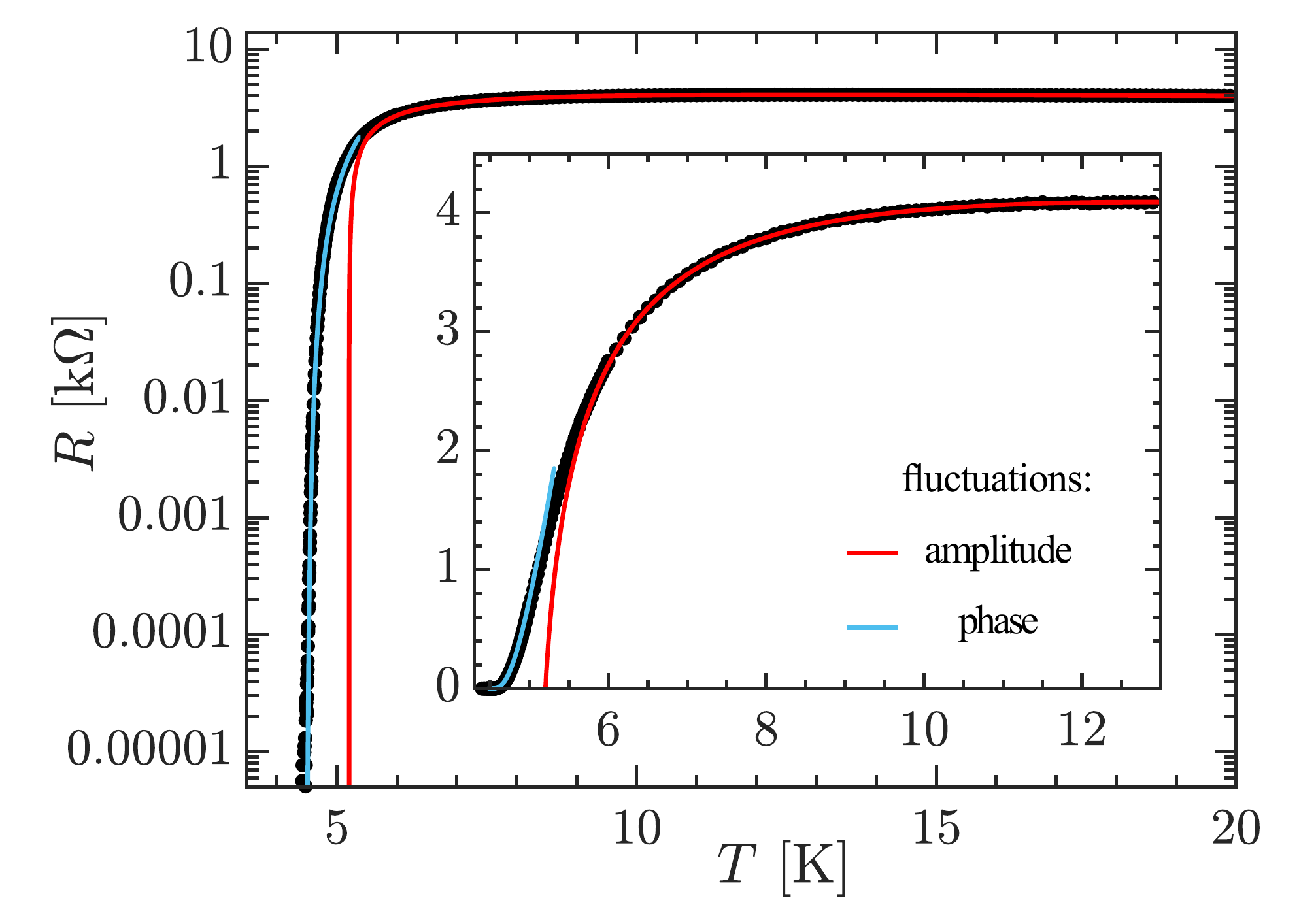}
	\caption[R(T)]{Sheet resistance, $R(T)$, as function of temperature of a 3.5~nm thick,  10~\textmu m wide and 2~mm long NbN meander on a logarithmic scale. In the normal state, we measure $R_N=4.093\,$k$\Omega$ at $15\,\text{K}$ and an electron density $n \simeq 4.5\cdot 10^{22}\mathrm{cm}^{-3}$ and $k_F\ell\simeq 2$ \cite{Supplement}. Red line: fit including amplitude fluctuations above mean field critical temperature $T_{c0}=5.203\,$\,K; blue line: fit to square-root-cusp expression corresponding to $T_\mathrm{BKT}=4.471\,$K and  $b=0.76$  (see text). Inset: Linear scale of $R$-axis emphasizes amplitude fluctuations.}
	\label{fig:R(T)}
\end{figure}

We start by establishing DC transport properties.  Figure \ref{fig:R(T)} shows resistance as function of temperature for a typical meander. The transition is strongly broadened by fluctuations of both amplitude and phase fluctuations of the order parameter \cite{Baturina_2012, Postolova2015, mironov2018,Kronfeldner2021, LarkinVarlamov2005, Supplement}.


Fitting $R(T)$ in Fig.~\ref{fig:R(T)}  for $R>0.6\,R_N$ (red line) reveals a mean-field transition temperature $T_{c0}=5.203$\,K (see \cite{Supplement} for details).  
Below $T_{c0}$, phase fluctuations of the order parameter generate resistance, where $R(T)$ has the 'square root cusp' form $R(T)\propto \exp(b/\sqrt{T/T_\mathrm{BKT}-1})$ \cite{AHNS1, HalperinNelson_1979, Baturina_2012, Postolova2015, mironov2018, Kronfeldner2021, Supplement}. Good agreement is found between theory (blue line) and experiment.  
Very similar results are found also for other devices with different width and length \cite{Supplement}.
\begin{figure}[t]
	\includegraphics[width=.48\textwidth]{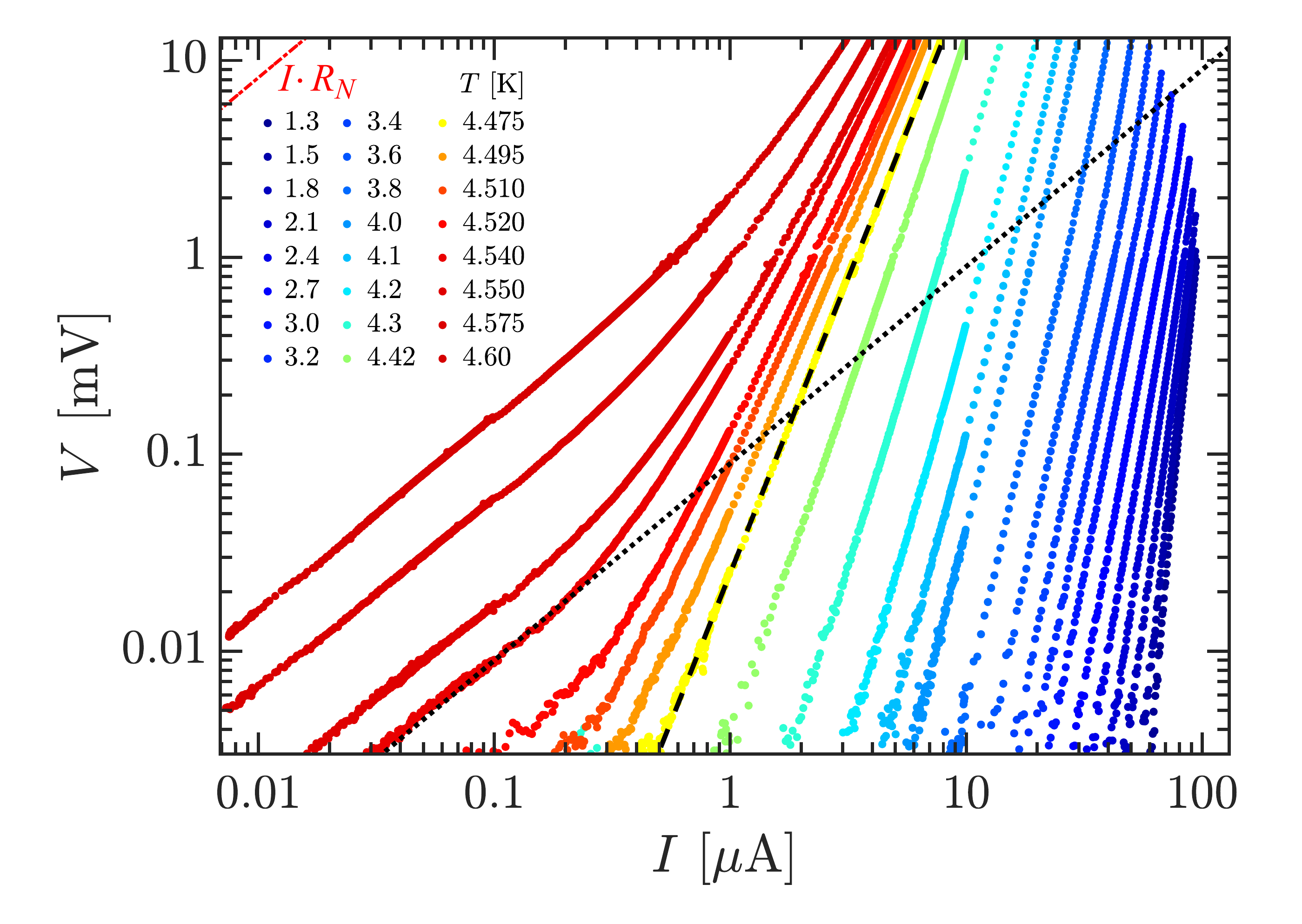}
	\caption[IV]{Voltage-current ($V(I)$) characteristics at different temperatures.  When collecting $V(I)$ over a wide range of current, we used fast sweeps (few sec), in order to avoid heating of the chips.  Straight lines in log-log display indicate power law behavior $V\propto I^{\alpha(T)}$, where $\alpha$ is related to $J_s(T)$ \cite{HalperinNelson_1979}. Dotted and dashed black lines corresponds to $\alpha=1$ and $3$, respectively. Red solid line in the upper left corner corresponds to $V=R_NI$ in the normal state. Slope $\alpha=3$ corresponds to $T=4.475$ (yellow). $J_s(T)$ extracted from the power law exponents $\alpha(T)$ is displayed in Fig.~\ref{fig:L(T)}. }
	\label{fig:IV}
\end{figure}
According to Halperin-Nelson (HN) theory, $V(I,T)$ takes the form \cite{HalperinNelson_1979} 
\begin{equation}\label{eq:HN-IV}
	V(I,T)=A(T)\cdot I^{\alpha(T)}
\end{equation}
with exponent $\alpha(T)={\pi J_s(T)}/{T}+1$ and prefactor $A(T)$. Hence, power-law behavior of $V(I)$-characteristics below $T_{\mathrm{BKT}}$, is another hallmark of the BKT-transition. Increasing temperature decreases $J_s$ and thus $\alpha$. At the universal transition point $\alpha=3$ (dashed in Fig.~\ref{fig:IV}) a characteristic jump to $\alpha=1$ is predicted.



\begin{figure}[t]
	\includegraphics[width=.48\textwidth]{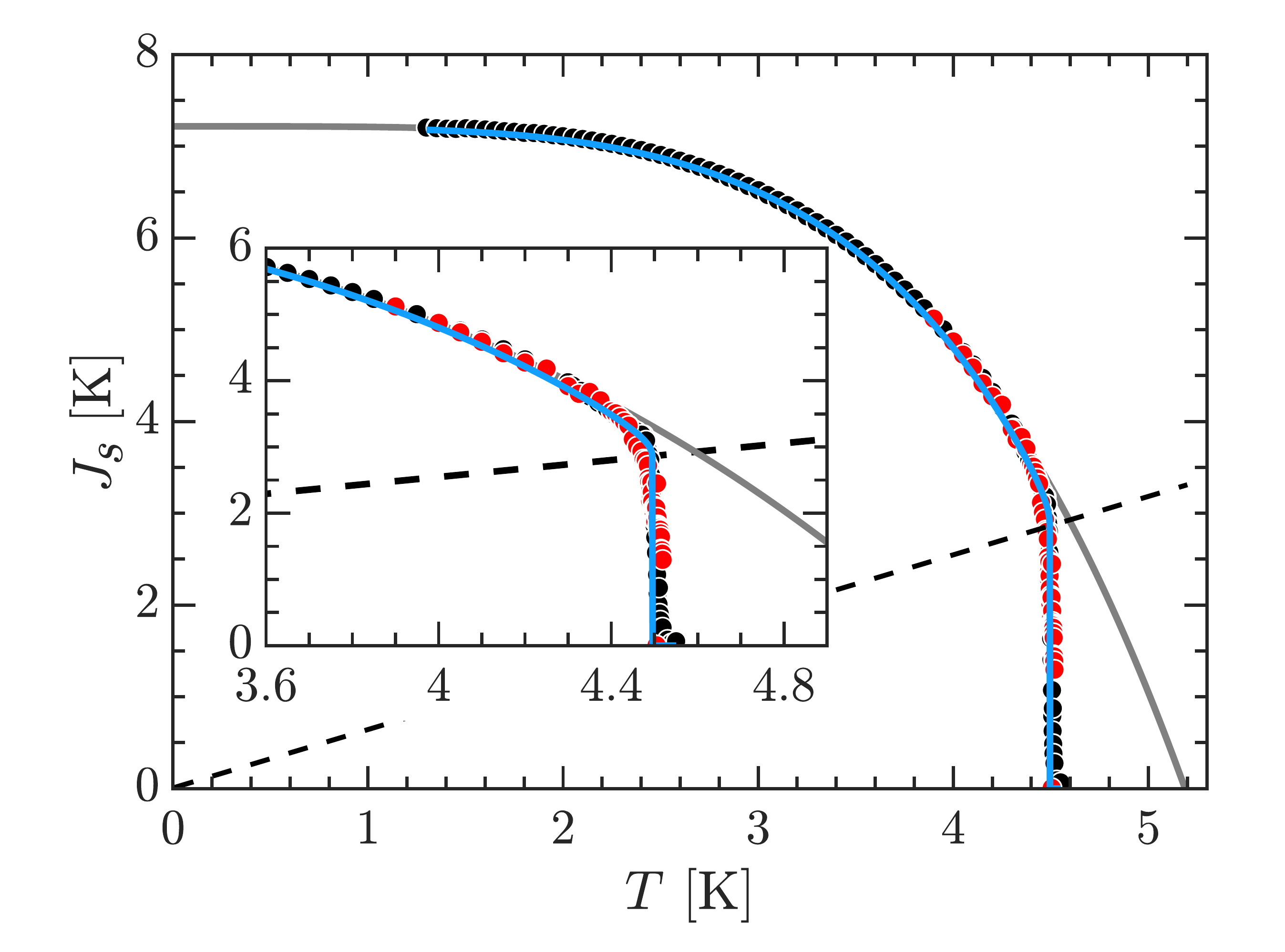}
	\caption[L(T)]{Superfluid stiffness, $J_s$, vs. temperature, $T$ for the same device as in Figs.~\ref{fig:R(T)} and \ref{fig:IV}. Black dots: $J_s$ extracted from kinetic inductance, red: $J_s$ extracted from nonlinear $IV$-characteristics (Fig.~\ref{fig:IV}), black: BCS-fit of low temperature part, light blue: Renormalization group calculation, dashed black: Nelson-Kosterlitz universal line. Inset: Zoom to the critical region near the jump. The fit parameters for the BCS fit are: $T_{c0}=5.175$\,K, $\Delta(0)/k_\mathrm{B} = 13.04$\,K and $J_s(0)=7.511\,$K. From $J_s(0)$ we extract $\lambda(0)=1.71\,$\textmu m (Eq.~\ref{eq:J_S}). The intersection of the data points with the universal line occurs at $T_{\mathrm{BKT}}= 4.488\,$K. The value of the vortex core energy extracted from the RG-fit is $\mu=19.0$\,K.}
	\label{fig:L(T)}
\end{figure}

In Figure~\ref{fig:IV} we present the evolution of $V(I)$ with temperature. At low temperatures and voltages the double-logarithmic plot reveals the expected power-law dependence. Above $T_\mathrm{BKT}$ and for sufficiently low current $V(I)$ is expected to be linear. The linear regime is limited first by current-induced dissociation of vortex-antivortex pairs, leading again to power-law behavior of $V(I)$, but now with values of $\alpha$ smaller than 3.  Note that the voltage level is orders of magnitude below $R_NI$ (red line in top left corner of Fig.~\ref{fig:IV}). 

At higher temperatures and in a wider voltage range $V(I)$ turns out to be much more complex  \cite{Baturina_2012,Postolova2015, Supplement}. At currents exceeding 10~\textmu A and $T\gtrsim T_\mathrm{BKT}$ heating effects start to play a role, rendering the $V(I)$-characteristics very complex and even dependent on the speed of current sweeps \cite{Supplement}. Above $T_{\mathrm{BKT}}$ the linear part of $V(I)$ may be buried in the background  noise, mimicking power-law behavior. Both above and below $T_{\mathrm{BKT}}$, heating effects can affect the observed power law exponent.  Based on  $V(I)$-characteristics \textit{alone}, it is thus very hard to judge whether values for $J_s(T)$ and even $T_{\mathrm{BKT}}$ are correct when extracted from $\alpha(T)$.  

As a consistency check,  $J_s(T)=T[\alpha(T)-1]/\pi$ extracted from $V(I)$ is plotted as red dots in Fig.~\ref{fig:L(T)} together with $J_s(T)$ (black dots) measured in equilibrium via the kinetic inductance $L_\square(T)$. The excellent agreement between the two independent data sets ensures the $\alpha(T)$ was extracted in the right regime and substantiates our analysis of the DC measurements. Very close to the universal transition point at $\pi J_s(T_{\mathrm{BKT}})=2T_{\mathrm{BKT}}$  (dashed line), $J_s(T)$ drops to zero within 50~mK.  The BKT transition is thus much sharper than in previous experiments on ultrathin NbN films \cite{Yong2013, Mondal_2011a, Mandal2020}. 
Also $T_{c0}$ and $T_{\mathrm{BKT}}$ obtained from $R(T)$ ($T>T_{\mathrm{BKT}}$) and $J_s(T)$ ($T<T_{\mathrm{BKT}}$) match within 1\% even though data were obtained in disjunct temperature intervals.

The gradual decrease of $J_s$ towards higher $T$ can be described by the BCS expression \cite{Yong2013,Mondal_2011b}
\begin{align}\label{eq:J_BCS}
	J_s(T) &= 
	J_s(0)\cdot\Delta(T)/\Delta(0)\cdot\tanh[\Delta(T)/(2k_\mathrm{B} T)]
\end{align}
(grey line), which accounts for the depletion of $J_s$ by quasiparticle excitations. In order to obtain a good match, it is established practice \cite{Yong2013,Mondal_2011b} to use $J_s(0)$, $\Delta(0)$ and $T_{c0}$ as independent fitting parameters \cite{Supplement}. 	The best fit is obtained for $T_{c0}=5.175$\,K, $J_s(0)=\beta J_\mathrm{BCS}(0)$ and $\Delta(0)=\gamma 1.764k_\mathrm{B}T_\mathrm{c0}$ with $\beta=0.7304$, $\gamma=1.432$. While $T_{c0}$  agrees within 30~mK or $0.5\%$ with the value obtained from the amplitude fluctuations of the order parameter (Fig.~\ref{fig:R(T)}), the ratio $\Delta(0)/k_\mathrm{B} T_\mathrm{c0}=2.521$ exceeds the BCS-value of 1.764 as observed earlier \cite{Yong2013,Semenov2009,Mondal_2011a,Mandal2020}. 

Moreover, $J_s(0)$ is smaller than the dirty limit BCS-prediction $J_\mathrm{BCS}(0)=\pi\hbar\Delta(0)/(4e^2k_\mathrm{B} R_N)$, consistent with the conjectured suppression of $J_s(0)$ by phase fluctuations \cite{Mondal_2011a}. 
The ratio $R_NJ_s(0)/T_\mathrm{c0}=5.940\pm0.3$~k$\Omega$ for several of our films with $R_N\simeq4$~k$\Omega$ agrees within a few percent with the BCS-value of $1.764\pi\hbar/(4e^2)=5.692$~k$\Omega$ \cite{Supplement}. This indicates that disorder effects in $J_s(0)/T_\mathrm{c0}$ are accounted for by $R_N$ alone, while both $J_s(0)$ and $T_\mathrm{c0}$ substantially differ from their dirty-limit BCS-expressions. An independent confirmation of the value of $\Delta(0)$ is highly desirable. 
Based on direct measurements of $\Delta(0)$ via tunneling spectroscopy, Carbillet et al.~proposed an interpretation of the large  $\Delta(0)/k_\mathrm{B} T_{c0}$ in terms of an underestimation of $T_\mathrm{c0}$ \cite{Carbillet2020}. In the latter work, $T_\mathrm{c0}$ was associated with the onset of the resistance, rather than $T_\mathrm{BKT}$. Here we can exclude this possibility, as our analysis allows for an unambiguous determination of $T_\mathrm{BKT}$ and $T_\mathrm{c0}$. 
%

We theoretically describe the  drop of $J_s(T)$, taking the BCS-fit to $J_s(T)$ as input for the BKT renormalization group (RG) equations \cite{Benfatto2009, maccari2020}. In this way,  data is closely reproduced by RG theory (blue line in Fig.~\ref{fig:L(T)}), assuming a vortex fugacity $\mu=19\,$K, or $\mu/J_s(0)\approx 2.5$, similar to values reported, e.g., in Ref.~\cite{Yong2013}.  It is instructive to compare $\mu$ with the loss of condensation energy $u_\mathrm{cond}$ in the vortex cores with effective radius $r_v$. We write $u_\mathrm{cond}=\mu/(\pi r_v^2d)\equiv B_c^2/2\mu_0=1/(2\mu_0)\cdot[\hbar/(2\sqrt{2}e\xi\lambda)]^2$, where $B_c$ is the thermodynamic critical field, and $\mu_0$ being the vacuum permeability. 
From the equation for $\mu$, we find $r_v/\xi(T_\mathrm{BKT})\simeq 2.2$.
Using the expression $T_\mathrm{BKT}=T_\mathrm{c0}(1-4\,Gi)$  with $Gi=7\,e^2\zeta(3)R_N/(\pi^3h)=0.0420$ being the Ginzburg-Levanyuk number \cite{Koenig2015}, we expect $T_\mathrm{BKT}^\mathrm{theo}=4.315$\,K, which is only 4\% smaller than $T_\mathrm{BKT}=4.488$\,K extracted from Fig.~\ref{fig:L(T)}.

\begin{figure}[t]
	\includegraphics[width=.45\textwidth]{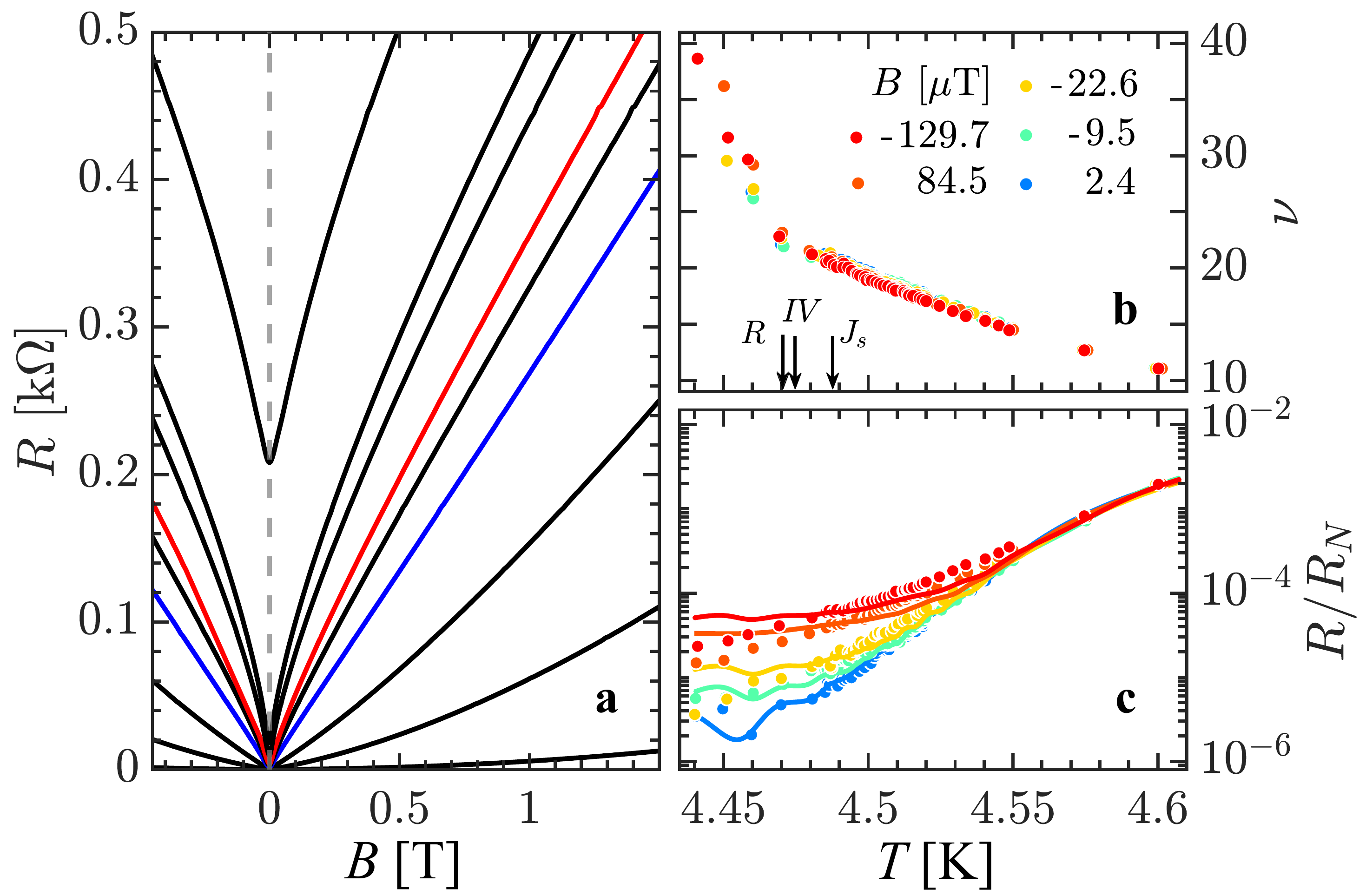}
	\caption[]{a) Magnetoresistance in the vicinity of $T_\mathrm{BKT}$ at different $T$. Curves correspond to temperatures (in K from top to bottom): 4.8, 4.6, 4.55, 4.4, 4.3, 4.0, 3.5, 2.5. Blue curve ($T=4.3$ K) shows linear slope, red curve ($T=4.45$ K) is closest to $T_\mathrm{BKT}$ from $J_s$. b) Low-field magnetoresistance expressed in term of the scaling parameter $\nu(T)$ (see text).  Arrows correspond to $T_\mathrm{BKT}$ from $R(T)$, $V(I)$ and $J_s(T)$, respectively. c) $R(T)$ at very low fields together with the scaling function (Eq.~\ref{eq:RB_scaling}). }
	\label{fig:R(B)}
\end{figure}

Finally, we investigate  signatures of the BKT-transition in magnetic field perpendicular to the film. In the high-field regime and near $T_\mathrm{BKT}$, $R(B)$ is expected to cross over from sublinear to superlinear behavior \cite{Garland1987}, signaling a transition from amplitude fluctuations of the order parameter to vortex pinning. In Fig.~\ref{fig:R(B)}a we  observe such 
cross-over at $T=4.3\,$K (blue line) slightly below the range  extracted from the other observables. This discrepancy is probably caused by lack of thermal cycling between curves. For the low-field regime, Minnhagen has derived the scaling law  \cite{Minnhagen1984}
\begin{equation}
	\label{eq:RB_scaling}
	\frac{B}{B_{c2}} =  \frac{R(B)}{R_N}\left[ 1-\left(\frac{1}{\nu}\ln\frac{R(0)}{R_N}\right)^2
	\left(\frac{R(B)}{R_N}\right)^{-2/\nu}
	\right]^{1/2}
\end{equation}
where $B_{c2}(T)=\Phi_0/2\pi\xi^2(T)$ is the upper critical field and the scaling parameter $\nu$ is a universal function of $T$ and $B$. In the Ginzburg-Landau (GL) limit, the coherence length can be written in the form $\xi(T)=\xi(T_\mathrm{BKT})[(T_\mathrm{c0}-T_\mathrm{BKT})/(T_\mathrm{c0}-T)]^{1/2}$. In Fig.~\ref{fig:R(B)}b we show separately measured low field data with field cooling procedure. From the measured $R(T,B)$ at fixed value of $B$, the function $\nu(T)$ can be determined from Eq.~\ref{eq:RB_scaling}, if $B_{c2}$ is given. Adjusting $B_{c2}(T_\mathrm{BKT})=6.1\pm1.7\,$\,T leads to a collapse of the set of $\nu(T)$-curves at low field (Fig.~\ref{fig:R(B)}b) that corresponds to a coherence length of $\xi(T_\mathrm{BKT})=6.7\pm0.6$\,nm. The error margins mark a deviation from optimal scaling by one dot size. This implies that $\nu(T,B)$ only weakly depends on $B$. The BKT-transition temperature is reflected as a cusp in  $\nu(T,B)$ which is located within 50~mK of $T_\mathrm{BKT}$ from the $R(T)$-curve (arrows in  Fig.~\ref{fig:R(B)}b).  

We use the scaling function $\nu(T)$ in order to predict $R$ vs.~$T$ at very low $B$ in  Fig~\ref{fig:R(B)}c.  Our directly measured $R(T)$ (dots) and data obtained from the scaling expression of Eq.~\ref{eq:RB_scaling} (lines) agree  well at the lowest $B$ as slightly less at higher $B$.
%

\vspace{1mm}
\emph{Discussion:} 
The central result of our work is the observation of a sharp, textbook-like BKT-transition in strongly disordered ultrathin NbN-films. Hence, the previously observed broadening \cite{Yong2013,Mondal_2011b} is no genuine consequence of strong (but homogeneous) disorder in superconducting thin films. 
A possible explanation for the sharpness is a more homogenous distribution of defects in our ALD-deposited films, as opposed to the sputter deposited films in earlier works \cite{Semenov2009,Carbillet2016,Carbillet2020}. Long-range correlated disorder can explain the observed broadening of the transition  in terms of a spatial variation of $J_s$ \cite{Benfatto2009, Benfatto_2013, Maccari2017}. 
%
On the other hand, short-range emergent granularity has been observed in STS for both sputter \cite{Carbillet2016, Kamlapure2013} and ALD \cite{Sacepe2008} deposited films alike. At least at the level of disorder in our films, intrinsic inhomogeneity in the gap distribution appears to be irrelevant at the large length scales that determine the BKT-transition.  This finding aligns with previous Monte Carlo simulations conducted on two-dimensional effective XY models \cite{Maccari2017, Maccari2018, Maccari2019}, which had highlighted that the mere presence of strong quenched disorder does not automatically lead to a broadening of the BKT transition. Instead, it is primarily the strong spatial correlation of the inhomogeneities that causes the smearing of the superfluid-stiffness jump at the critical point.

Most often, the presence of a BKT-transition is deduced from the non-linearity of $V(I)$-characteristics. However, this is not straightforward, as $V(I)$ often is strongly affected by heating phenomena. First, power-law behavior can also occur slightly above $T_\mathrm{BKT}$, where only relatively few vortex-anti-vortex pairs are dissociated. A linear regime exist at the lowest currents only, while already at current densities $\lesssim100\,$A/m$^2$ current-induced dissociation dominates over thermal dissociation, leading to power law behavior with $\alpha < 3$ that are not considered by standard theory.

These observations are important, because a substantial fraction of the recent literature on ultra-thin materials analyzes $IV$-characteristics in the high-power regime above $T_{\mathrm{BKT}}$ in terms of BKT-behavior (see e.g.~\cite{Reyren2009,Zhang_2014,Lu_2015,cao_super_2018,Park_2021,Zhou_Young_2021, Enze_Zhang2023}). Our work shows that power-law exponents obtained in this regime are unrelated to BKT-physics. 

\emph{Conclusions:} We have shown that ultra-thin superconducting films with strong, but homogeneous, disorder feature a sharp BKT-transition without significant broadening. All relevant observables display quantitatively consistent results, allowing for a precise determination of the BKT- and mean-field transition temperatures as well as other parameters governing the films. Our study lays the ground for future controlled studies of the statics and dynamics of the BKT transition in ultra-thin superconductors when approaching to the superconductor-insulator transition.    

\begin{acknowledgments}
	We would like to thank M.~Ziegler and V.~Ripka for NbN film deposition by ALD in the Leibniz IPHT clean room and T.~Baturina, E.~König, I.~Gornyi, A.~Mirlin,  P.~Raychaudhuri, A.~Ghosal and F.~Evers for helpful comments. The work was financially supported by the European Union’s Horizon 2020 Research
	and Innovation Program under grant agreements No 862660 QUANTUM E-LEAPS.
\end{acknowledgments}



%

	\let\oldaddcontentsline\addcontentsline
	\renewcommand{\addcontentsline}[5]{}
	%
	\let\addcontentsline\oldaddcontentsline
	

\clearpage
\newpage
\beginsupplement

\begin{center}
	{{\bf\Large SUPPLEMENTARY MATERIAL}\\
		Sharpness of the Berezinskii-Kosterlitz-Thouless transition in ultrathin NbN films\\
		A. Weitzel {\it et al.}\\
		Institut für Experimentelle und Angewandte Physik, University of Regensburg, Regensburg, Germany
	}
\end{center}

\tableofcontents

\begin{figure}[t]
	\includegraphics[width=.35\textwidth]{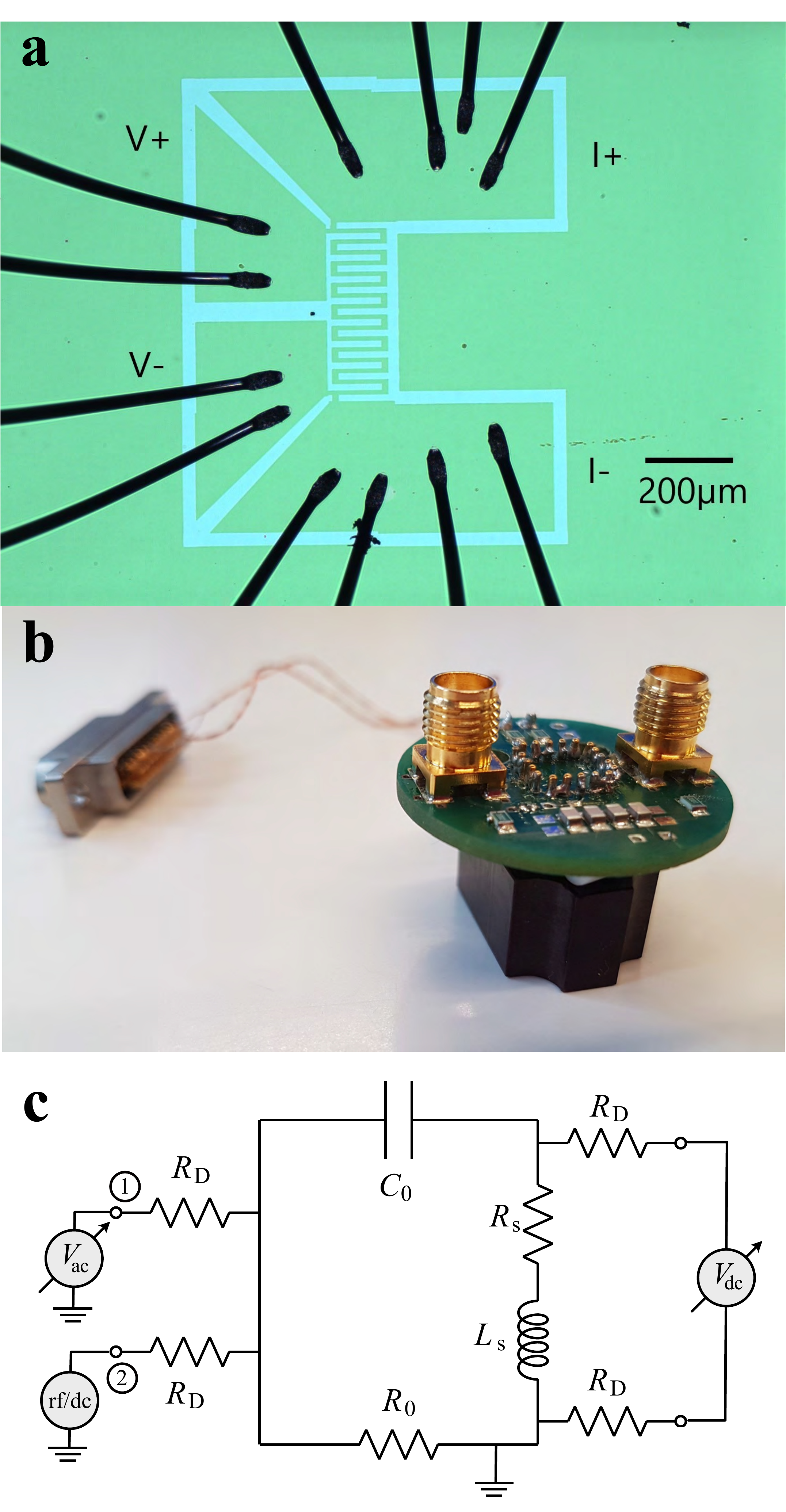}
	\caption[]{(a) Photograph of the sample discussed in the main text. Green is NbN, blue SiO$_2$/Si. (b)  Photograph of a typical RLC circuit. The brown block made from Tecasinth is the sample block for loading LCC20 chip carriers. (c) Circuit diagram used for measurements presented in the main text. Input and output of the VNA are connect to ports 1 and 2, the DC current source to port 1.}
	\label{fig:sup:RLC}
\end{figure}

\section{Materials and Methods}

We fabricate our samples from ultra-thin  NbN films grown by atomic layer deposition (ALD) on top of an amorphous, about 550 nm thick SiO$_2$ layer onto silicon substrate (thermally oxidized silicon wafer with [100] orientation). 
The growth conditions are completely different for ALD and sputtering. The average growth rate is more than one order of magnitude higher in sputtering. Furthermore, the substrate temperatures are about 300 K higher as well as the kinetic energy of the impinging particles (atoms, ions, electrons). The combination leads to a pronounced growth of NbN crystallites with different orientations linked to each other via grain boundaries \cite{Semenov2009,Carbillet2016,Carbillet2020} In contrast, the  layer-by-layer growth of ALD in combination with a low atom mobility on the substrate surface leads to high structural (lateral) homogeneity. Thus, a high density of point defects is accompanied by less pronounced grain boundary formation. 
A detailed comparison of the structure for ALD-grown and sputtered ultrathin ($\lesssim$5~nm)  NbN films has not been reported yet and may warrant further study. We believe that such large-scale extrinsic inhomogeneity is much weaker in our ALD-grown films, as compared to epitaxial films. 

A typical sample chip with a 10~\textmu m wide meander and bond wires is shown in Fig.~\ref{fig:sup:RLC}a.

\begin{figure}[t]
	\includegraphics[width=.5\textwidth]{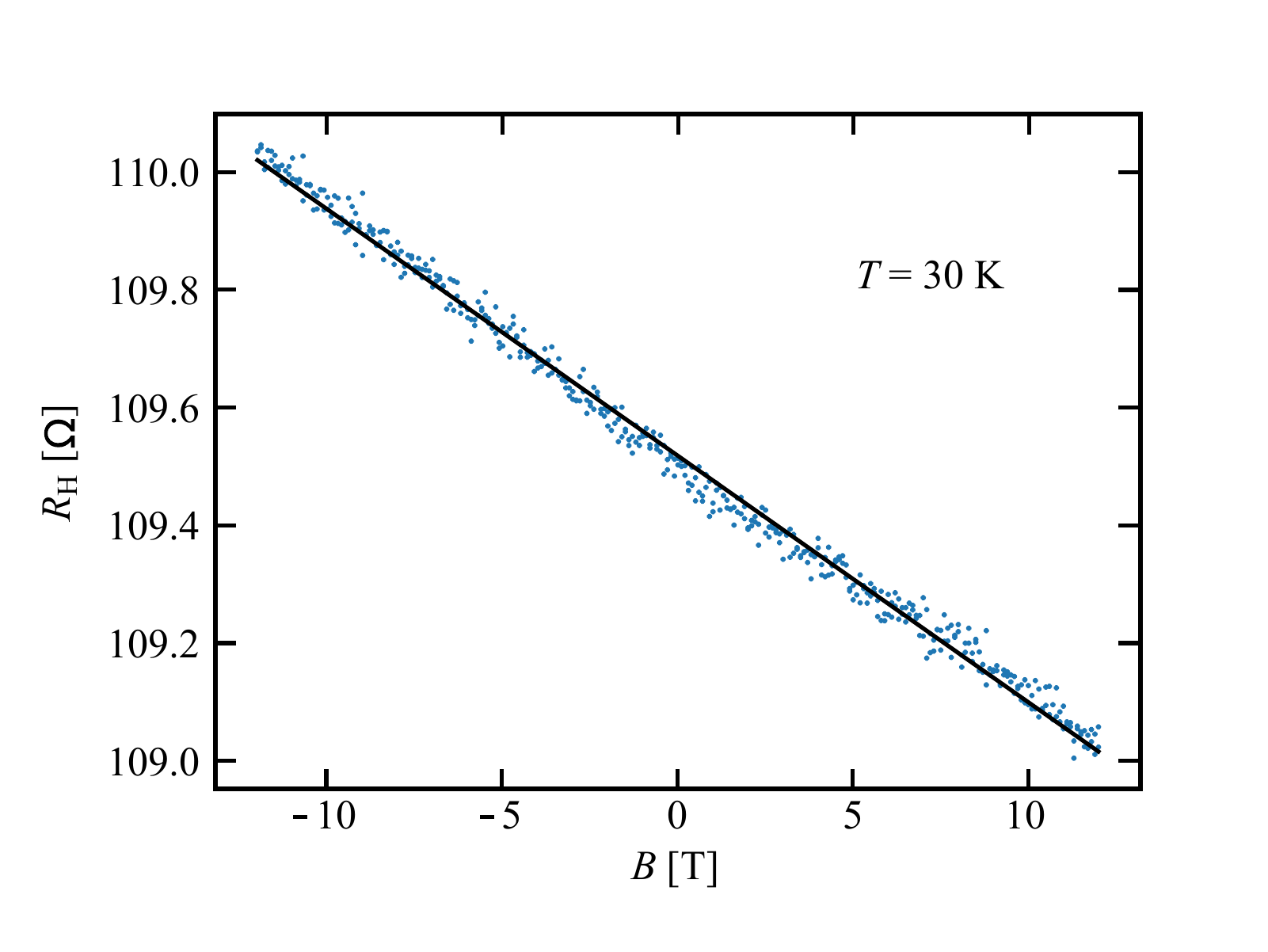}
	\caption[]{Hall resistance $R_\mathrm{xy}$ measured on a Hall bar  patterned into another chip from the same wafer used for the experiments in the main text.}
	\label{fig:sup:Hallmessung_30K}
\end{figure}

$R(T)$ presented in the main text is measured in 4-contact geometry with a standard voltage source (Yokogawa GS 200) and Nanovoltmeter (Agilent 34420A). Each data point corresponds to an $V(I)$ characteristic from which the zero bias resistance is inferred by linear fit. The current is swept in a small range from -10 to +10 nA to minimize heating effects. On a Hall bar with thickness $d=3.5$~nm,  width $w=40$~\textmu m  and length $l=400$~\textmu m  at  $T=30$~K at 3.929~Hz we have measured the Hall voltage shown in Fig.~\ref{fig:sup:Hallmessung_30K} and found a sheet resistance of $R_N\ =\ 4.014$~k$\Omega$ and a carrier density $n\ =\ 4.5\cdot10^{28}~\text{m}^{-3}$.

The chip is then integrated into a RLC-circuit mounted at low temperatures (Fig.~\ref{fig:sup:RLC}b,c). The meander inductance $L(T)$ is obtained from resonance curves measured with a vector network analyzer (VNA, Rohde und Schwarz ZNL3) coupled capacitively to high frequency lines in the cryostat. The input power $P=80$~dBm is chosen  such that the quality factor of the circuit remains unchanged upon further reduction. We amplify the output signal at room temperature by 56 dB using a Miteq AU 1447 amplifier.
Fast $V(I)$ characteristics are measured in 4-contact with a FPGA-base source-measure unit (Nanonis Tramea) with sweep duration of 1-7 s. We measure low voltage regime by using a Femto DLPVA with amplification 80 dB and bandwidth 100 kHz. In parallel, we measure $V(I)$ in the high voltage regime without amplification. Low $V$ and high $V$ regimes overlap over roughly an order of magnitude. All lines are filtered at room temperature by $\pi$ filters with cutoff frequency 100 MHz.

\begin{figure}[t]
	\includegraphics[width=.5\textwidth]{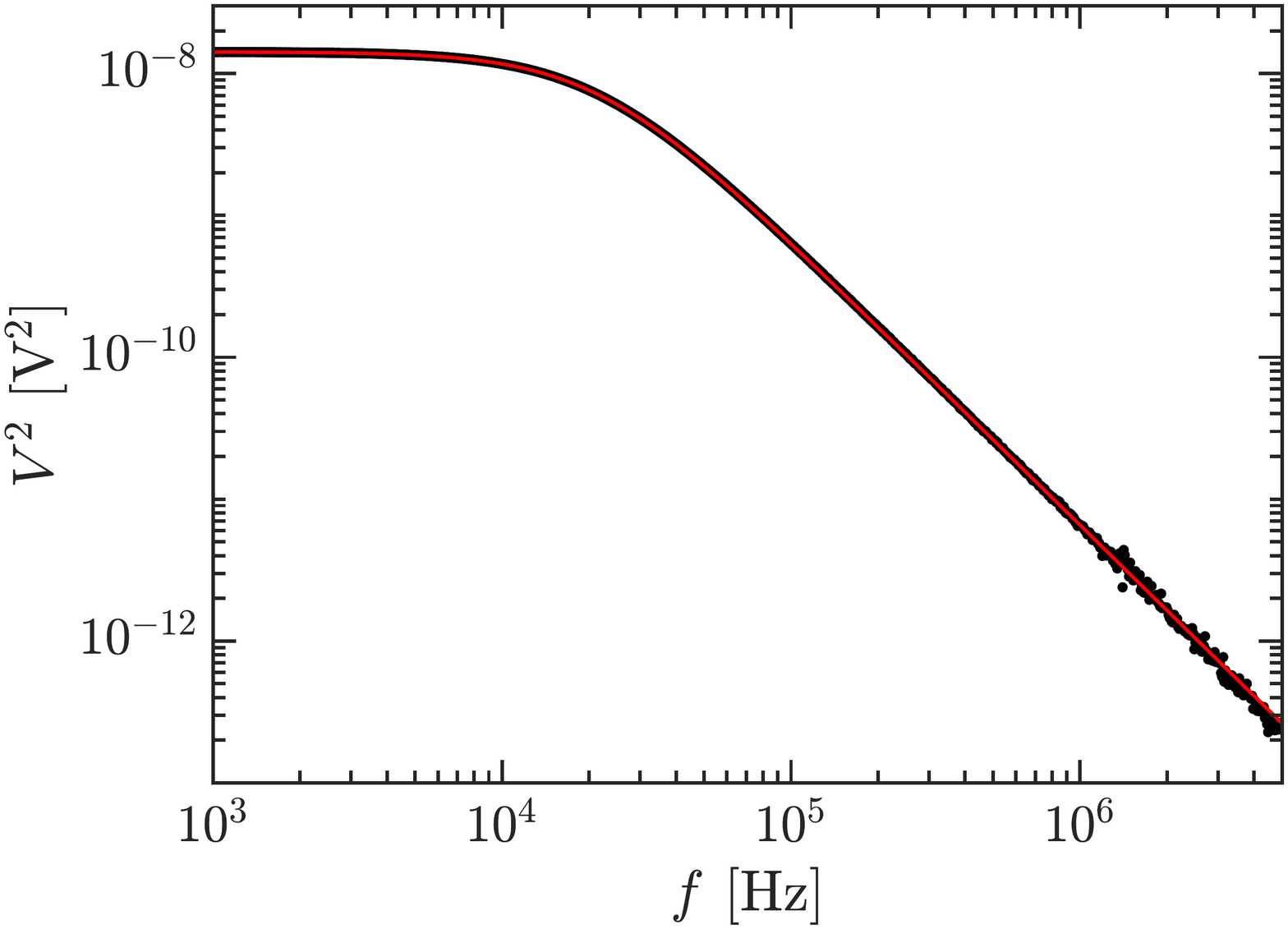}
	\caption[]{$V^2(f)$ measured above $T_c$ (black dots), together with best fit according to Eqn. \ref{eq:C0} (red line) to determine the capacitance $C_0$ of the LC-circuit. Grey shaded area indicates region used to perform the fit. The downturn at frequencies $\gtrsim 10\,$MHz results from stray capacitance between the pogo pins.}
	\label{fig:sup:C0}
\end{figure}

\section{Circuit Design}
\begin{figure}[b]
	\includegraphics[width=.45\textwidth]{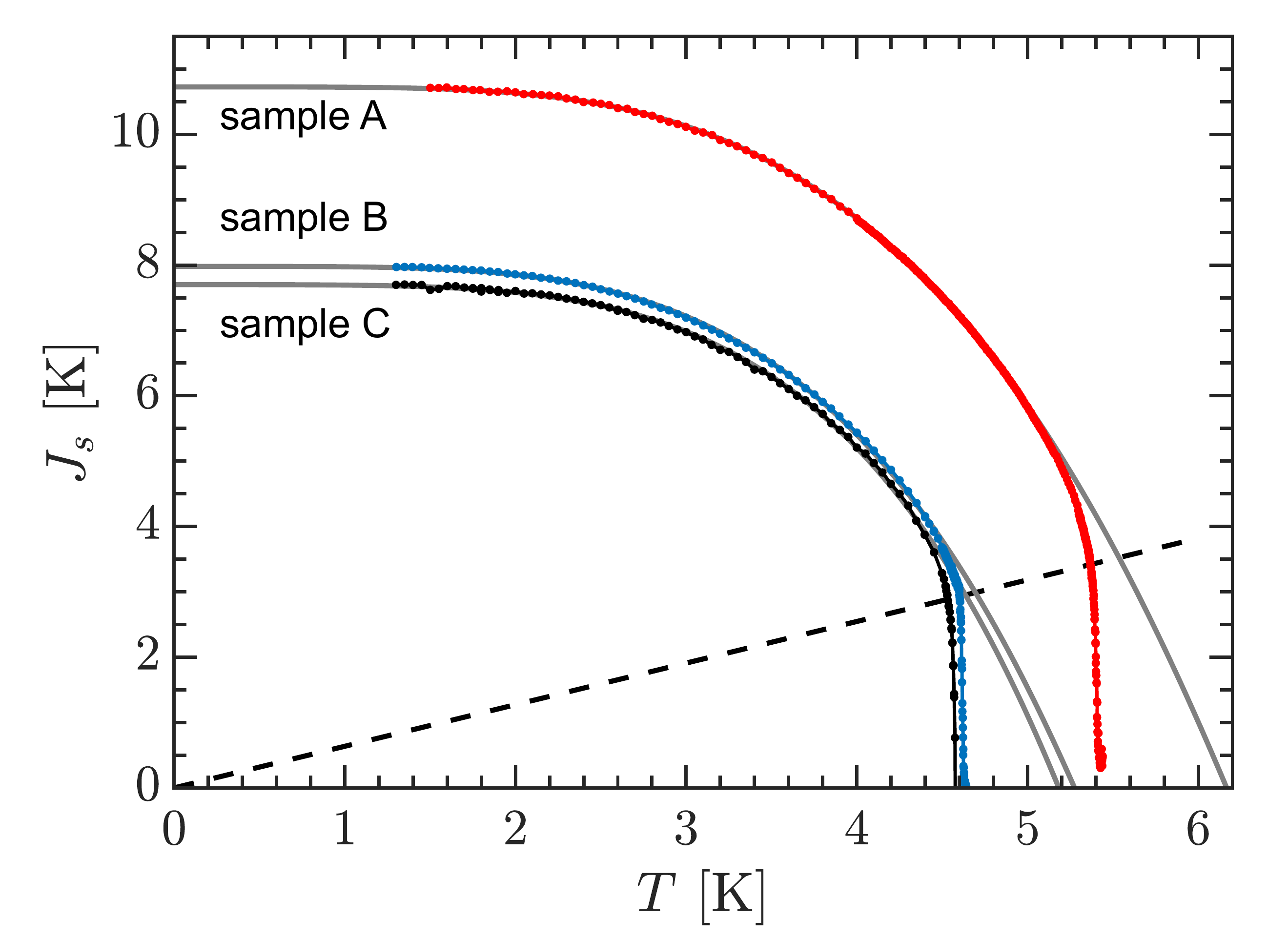}
	\caption[200um]{Superfluid stiffness of NbN meanders with  200 squares and 10 \textmu m width (samples A,B) and 82~squares and 200~\textmu m width (sample C),  respectively. The black dashed line is the universal BKT transition line. The differences in $T_\mathrm{c0}$ and $J_s(0)$ are caused by different levels of oxidation of the films in air that occurs on the scale of several months. Sample B and the sample in the main text refer to the same meander that was measured twice with a time delay of several weeks.}
	\label{fig:J_s_200um}
\end{figure}

The sample chip is placed in a LCC20 chip carrier and is then mounted into a cold RLC circuit. Figure~\ref{fig:sup:RLC}b shows a photo of a typical set-up, with green PCB-board as well as brown sample holder beneath. The sample holder is made from Tecasinth (polyimide) and holds an array of pogo pins that establish electric and thermal contact to standard LCC20 chip carriers, and via Al bond wires also to the sample on a Si-chip.  To drive and read-out the circuit coaxial cables are needed, which are connected via the SMA-ports on top of the PCB. The other two ports are used to measure the voltage drop over the sample.  
Fig.~\ref{fig:sup:RLC}c shows the corresponding circuit diagram. Four resistors $R_D$ are added to decouple the resonator from the environment, i.e. the lead impedances. For the sample discussed in the main text, a SMD capacitor with capacitance $C_0$ is placed in parallel to the sample with kinetic inductance $L_s$ and resistance $R_s$, forming the resonator. In the fully superconducting state, a residual resistance $R_0\simeq90$~m$\Omega$ limits the internal quality factor. The lead inductance $L_0=19.8$~nH adds to the sample inductance $L_s$, resulting in a total inductance $L_\mathrm{tot}=L_s+L_0$. In early measurements (Fig.~\ref{fig:J_s_200um}) we had placed an additional copper coil in series with the sample adding a typical inductance of 300~nH and a resistance of 50~m$\Omega$.

Using the transmission matrix formalism, the element $S_{21}$ of the transmission matrix between terminal 1 and 2 (Fig.~\ref{fig:sup:RLC}) can be calculated:
\begin{equation}
	S_{21}(\omega)= \frac{2Z_lQ}{C_0R_x^2}\frac{1}{\omega_0+2iQ{(\omega-\omega_0)}}\;.
\end{equation}
Therefore, the measured signal can be described by
\begin{equation}
	V^2(f)=A\cdot\bigg|\frac{Z_lQ}{\pi C_0R_x^2}\frac{1}{f_0+2iQ({f-f_0})}\bigg|^2\;,
	\label{eq:sup:V^2(f)}
\end{equation}
where $V$ is the voltage at the input of the network analyzer, $f$ is the frequency, $A$ is a scaling parameter containing the drive level, $Q$ is the quality factor, $R_x=R_D+Z_l$ with $R_D=992\,\Omega$ being decoupling resistors that separate the circuit from lead impedances and $Z_l=50\Omega$ is the impedance of the cables.
Changes of resonance frequency can be directly connected to changes of the sample's kinetic inductance.



\section{Circuit Calibration}
We determine $C_0$, the capacitance, by heating the circuit with sample to a temperature above $T_c$, such that the resistance of the sample is very large ($0.8$~ M$\Omega$) when compared to $\omega (L_\mathrm{tot})$. In this regime, the circuit can be effectively modeled as a low-pass filter, with transfer matrix element:
\begin{equation}
	V^2(f)=A\cdot\bigg|\frac{Z_l}{R_x+i\pi C_0 fR_x^2}\bigg|^2~\;.
	\label{eq:C0}
\end{equation}
Figure~\ref{fig:sup:C0} shows measured $V^2(f)$ together with the best fit according to Eq.~\ref{eq:C0}, giving $C_0=10.34$ nF, close to the nominal value of the capacitors ($10$~nF).  

\begin{figure}[t]
	\includegraphics[width=.5\textwidth]{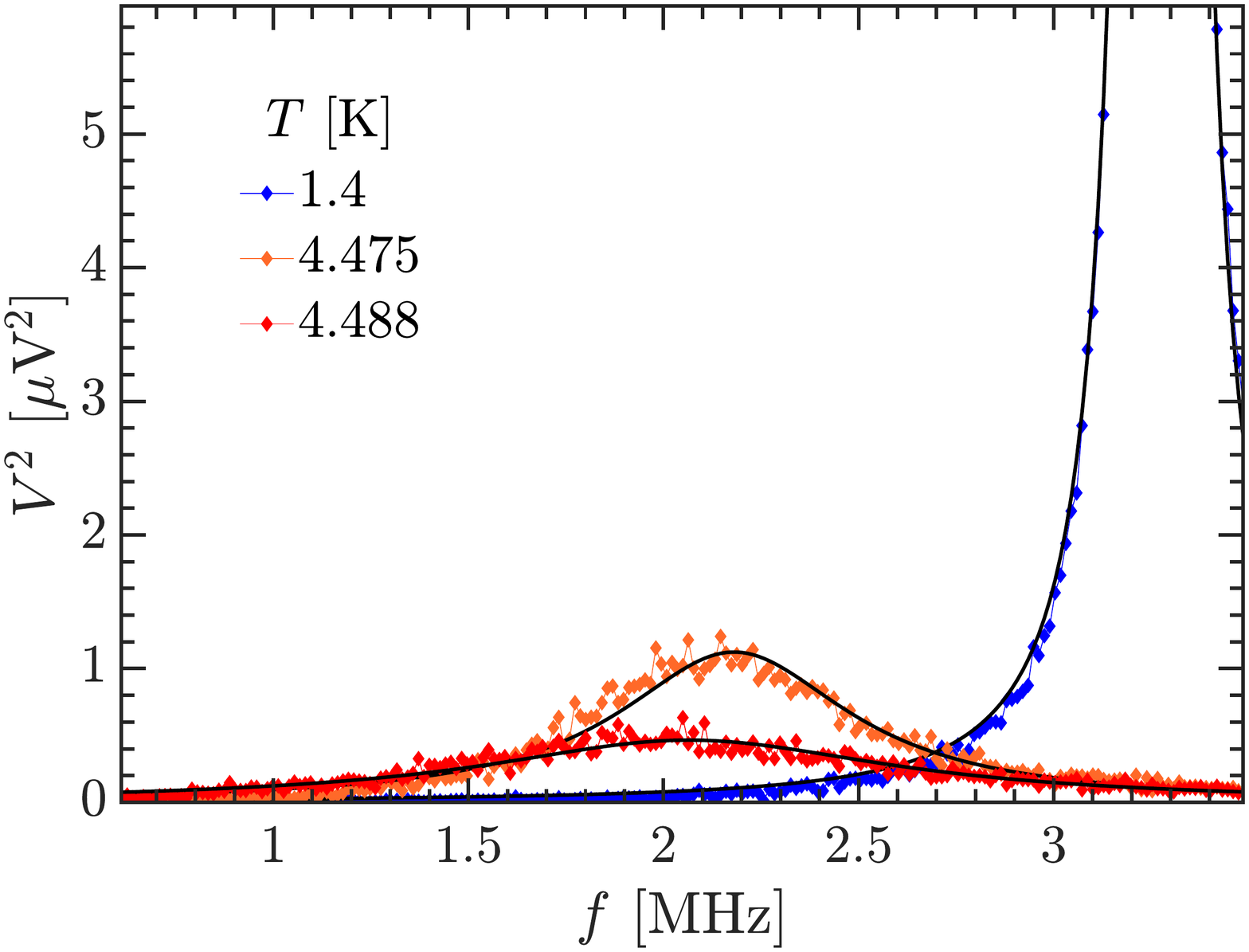}
	\caption[]{Three exemplary spectra, far away from $T_\mathrm{BKT}$ (blue), close to (orange) and at $T_\mathrm{BKT}$ (red). Solid lines are fits to Eq.~\ref{eq:sup:V^2(f)}. Corresponding $Q$-factors are 35 (blue), 3 (orange), 1.5 (red).}
	\label{fig:sup:spectra}
\end{figure}

For resonators with an additional inductor in series with the sample, we determine its inductance $L_0$ by replacing the sample with a straight bond wire of negligible resistance (50~m$\Omega$) and inductance ($\simeq1$~nH). From the resulting resonance frequency $f_{00}$, we  determine 
\begin{equation}
	L_0=\frac{1}{(2\pi f_{00})^2C_0}
	\label{eq:L0}
\end{equation}
and find typical values around 220~nH.

\section{Exemplary Spectra}

\begin{figure}[t]
	\includegraphics[width=.5\textwidth]{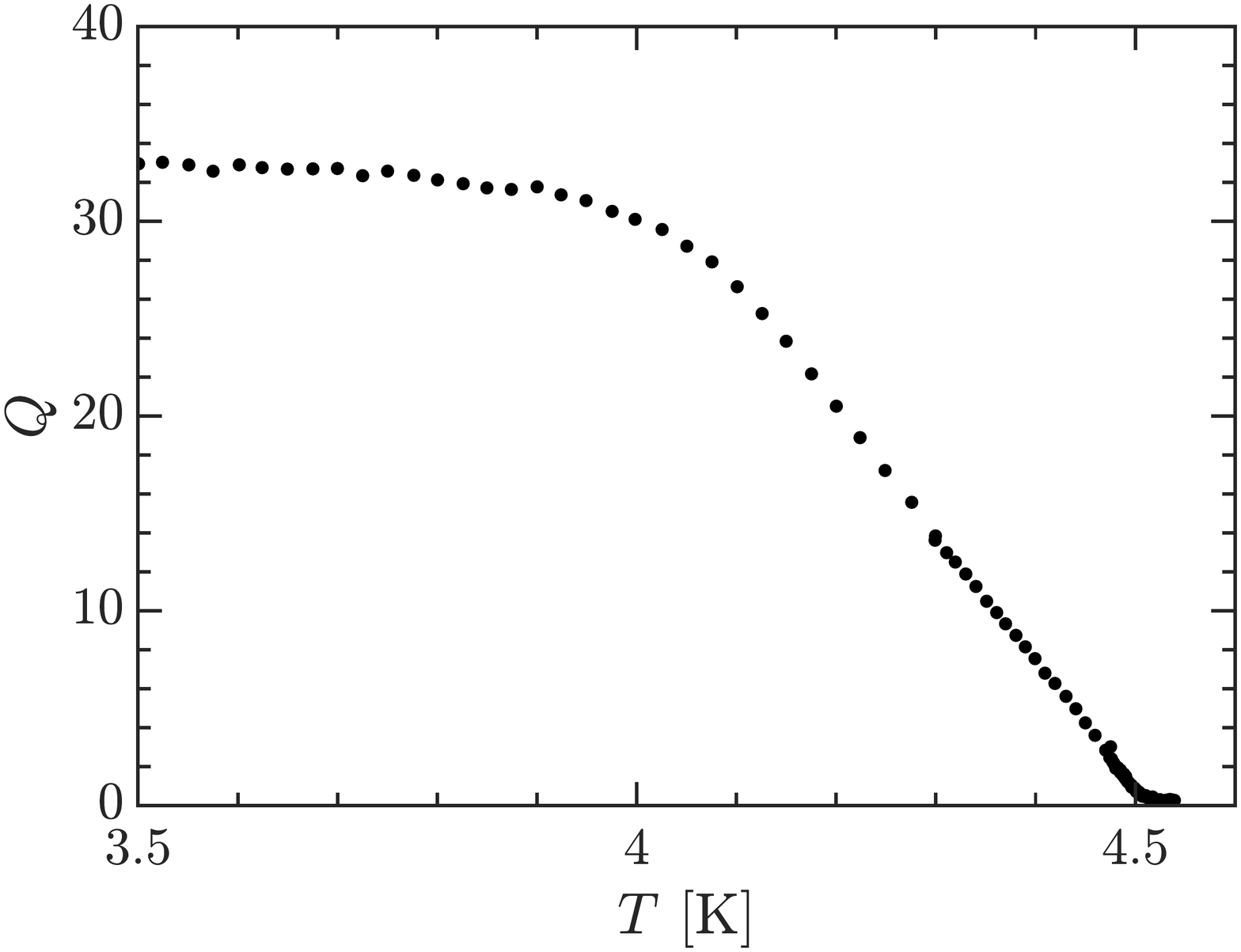}
	\caption[]{Quality factor  as function of temperature close to $T_\mathrm{BKT}$.}
	\label{fig:sup:Q(T)}
\end{figure}

We present some exemplary spectra in Fig.~\ref{fig:sup:spectra}. At $T_\mathrm{BKT}=4.488$ K a resonance is still clearly resolvable, albeit with a strongly suppressed $Q$ factor compared to low temperatures.  $Q(T)$ close to $T_\mathrm{BKT}$ is shown Fig.~\ref{fig:sup:Q(T)}. About 0.5~K below $T_\mathrm{BKT}$, $Q(T)$ decreases rather linearly towards $T_\mathrm{BKT}$. A possible explanation for the decrease is that the oscillatory motion of the increasing number of thermally excited, but still bound, vortex-antivortex pairs around their equilibrium distance gives rise to an additional dissipation channel.

\section{Quality Factor}
The quality factor of a resonance curve is defined as the ratio of resonance frequency over the full width at half maximum (FWHM). For the circuit displayed in Fig.~\ref{fig:sup:RLC}, $Q$ can be written as the combination of the circuit's internal and external quality factors $Q_e$ and $Q_i$:
\begin{equation}
	Q=\frac{f_0}{\Delta f_\mathrm{FWHM}}
	=\bigg[ \frac{1}{Q_i}+\frac{1}{Q_e} \bigg]^{-1}\;.
\end{equation}
where 
\begin{equation}
	Q_i = \frac{1}{2\pi f_0C_0 R_s}\quad \text{and}\quad Q_e=\frac{\omega_0C_0(R_D+Z_l)}{2}\;.
\end{equation}
%
%
If $C_0$ as well as $Z_l$ and  $R_0$ are known, small sample resistances $R_s$ can be determined from the internal quality factor.

\begin{figure}[t]
	\includegraphics[width=.5\textwidth]{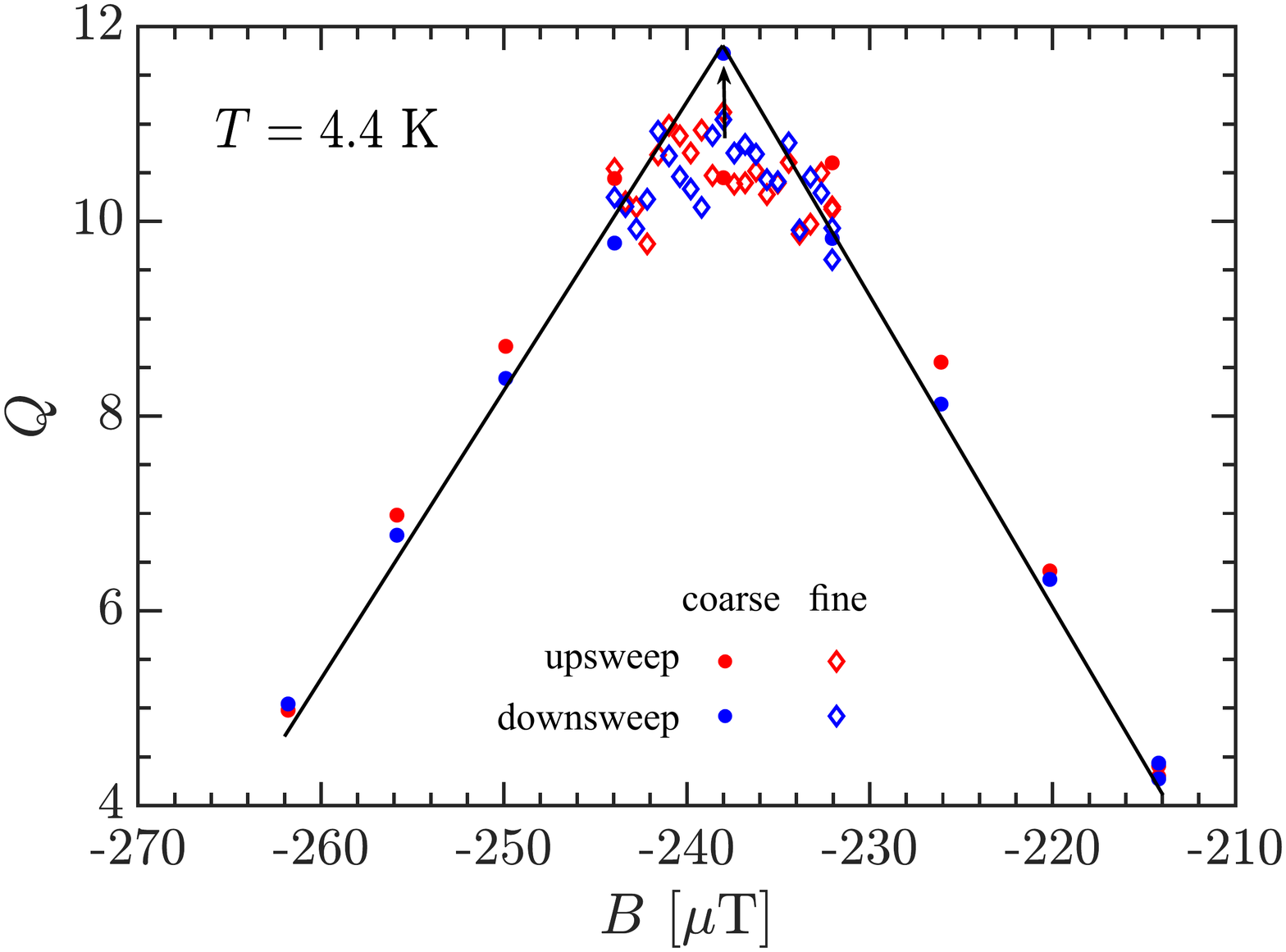}
	\caption[]{Quality factor  as function of magnetic field for the device discussed in the main text. Solid lines are guides to the eye. Arrow indicates the optimal compensation field as determined from the data.}
	\label{fig:sup:Q(B)}
\end{figure}

\section{Magnetic Field Compensation}
By optimization of the quality factor $Q(B)$, the perpendicular magnetic field can be compensated down to a few \textmu T. In Fig.~\ref{fig:sup:Q(B)} $Q(B)$ shows a sharp maximum with a width of $\simeq~10\,$\textmu T. The precise location of the maximum is obtained from the linear extrapolation of $Q(B)$ from both sides of the maximum. Negligible hysteresis was observed for  up- and down-sweeps. For the experiments presented in the main text, we chose $B_{\mathrm{comp}}=238$~\textmu T to compensate 
residual fields, indicated by an arrow in Fig.~\ref{fig:sup:Q(B)}. Experience shows that after heating the solenoid above $T_c$  strongly improves the stability of the residual field such that compensation is reliable over several days. All experiments in zero field presented in the main text were performed within six days after field compensation.

\section{Fluctuation Contributions to $R(T)$ }
\label{sec:qcc}
\begin{figure}[t]
	\includegraphics[width=.5\textwidth]{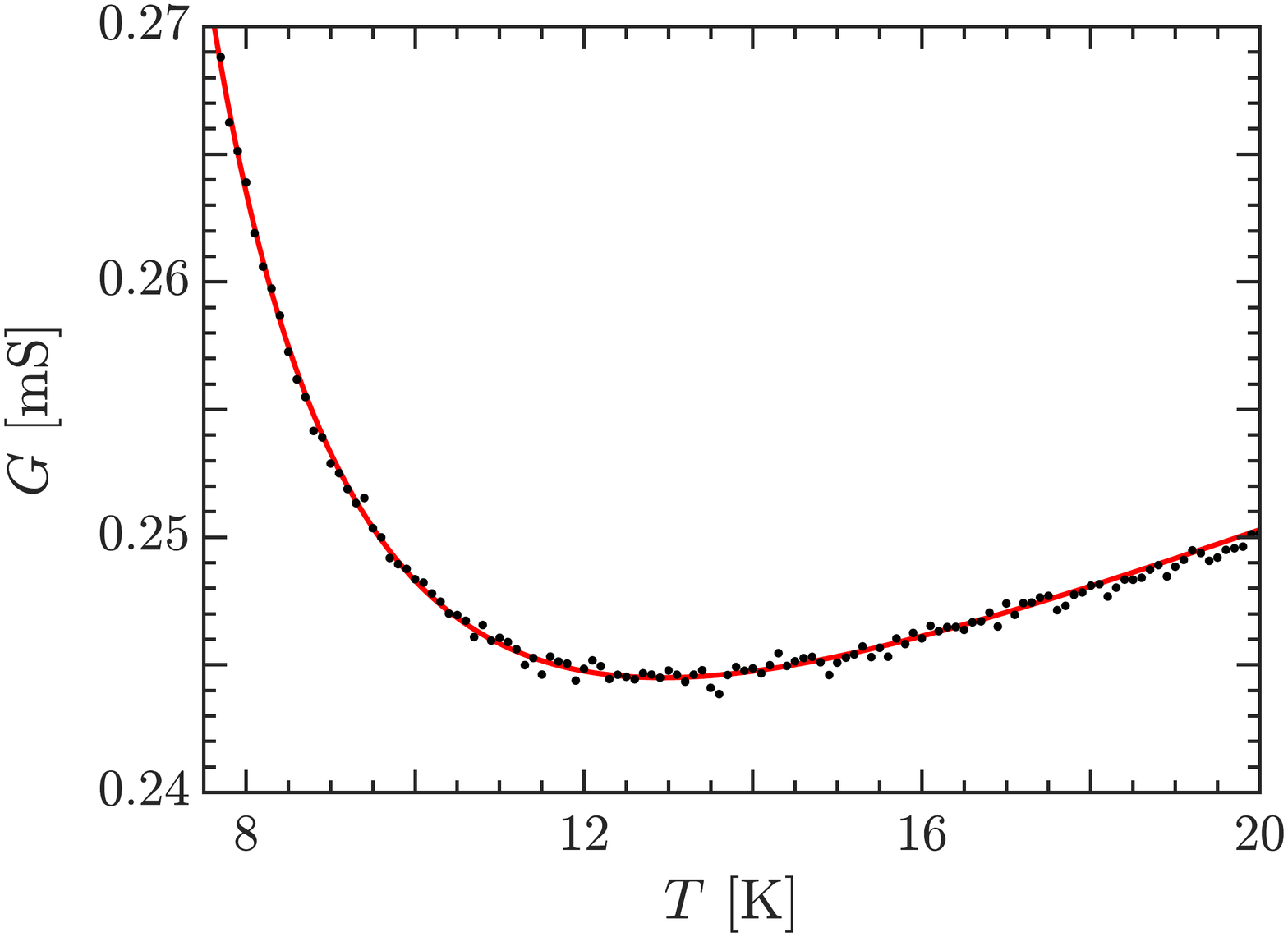}
	\caption[]{Sheet conductance $G(T)$ above $T_\mathrm{c0}$ together with the best fit according to Eq.~\ref{eq:G(T)}. }
	\label{fig:sup:G(T)}
\end{figure}

Our fit of the intrinsically broadened $R(T)$-curve consists of several contributions (see \cite{LarkinVarlamov2005,Baturina_2012,Postolova2015} and the references therein):
\begin{equation}\label{eq:G(T)}
G(T)=\frac{1}{R_\mathrm{Drude}} +\Delta G_{AL} +\Delta G_{MT} +\Delta G_{WL} +\Delta G_{ID} 
\end{equation}
Besides the normal-state resistance  $R_N$, the Aslamazov-Larkin (AL) contribution $\Delta G_{AL}$ describes the fluctuation of Cooper pairs above $T_\mathrm{c0}$
\begin{align}
	\Delta G_{AL} &= G_{00}\cdot 
	\frac{\pi^2}{8\,\ln(T/T_\mathrm{c0})}\;.
\end{align}
Fluctuating Cooper pairs also affect the diffusion coefficient of unpaired electrons due to the interaction with the Cooper pairs. This included as $\Delta G_{MT}$, named after Maki and Thompson
\begin{align}
	\Delta G_{MT} &= G_{00}\cdot
	\beta\big( T/T_\mathrm{c0}, \delta\big)\cdot
	\ln{ \left( \frac{\ln(T/T_\mathrm{c0})}{\delta} \right) }\;,
\end{align}
where $G_{00}$ is the conductance quantum, while the Larkin function $\beta$ and the Maki-Thompson pair breaking parameter $\delta$ are given by
\begin{align}
	\delta &= \frac{\pi \hbar}{8k_\mathrm{B} T \tau_\phi}\qquad\mathrm{and,}\\
	\beta\big( T/T_\mathrm{c0}, \delta\big) &\approx \frac{\pi^2}{4}\ \frac{1}{\ln(T/T_\mathrm{c0})-\delta}
	\label{eq:supp:beta}
\end{align}
where the approximate form of $\beta(T)$ in Eqn.~\ref{eq:supp:beta} is valid in the limit $\ln(T/T_\mathrm{c0})\ll 1$ \cite{SantosAbrahams1985}.
Finally, the normal state weak localization and interaction corrections are responsible for the resistance maximum above $T_\mathrm{c0}$ and read:
\begin{align}
	\Delta G_{WL}+\Delta G_{ID} &= G_{00}\cdot
	A \cdot \ln \left( \frac{k_\mathrm{B} T\tau}{\hbar} \right) \;.
\end{align}
Four parameters are determined from a curve fit according to Eq.~\ref{eq:G(T)}:  $T_\mathrm{c0}$, $A$, $\delta$ and $R_\mathrm{Drude}$.   The Drude-scattering time $\tau$ enters only logarithmically and is tied to the Drude resistance via Eqs.~\ref{eq:kF_kFl} below. 
These parameters are mainly determined by different features of the curve in Fig.~\ref{fig:sup:G(T)} : $T_\mathrm{c0}$ by the sharp rise in the low $T$-region, $A$ by the slope above the minimum, $\delta$ by the curvature around the minimum and $R_\mathrm{Drude}$ by the value of $G$ at the minimum. 
Nevertheless, three  of the parameters ($R_\mathrm{Drude}, A, \delta$) show some mutual dependency which limits their accuracy to $\simeq 10\%$.

 The combined strength of $\Delta G_{WL} +\Delta G_{ID}$ is measured by the prefactor $A=3.9$. For the pair-breaking parameter we find $\delta\simeq0.42$.
Due to the divergence of the fluctuation term $\Delta G_{AL} +\Delta G_{MT}$ at the mean field transition temperature, $T_\mathrm{c0}$ can be determined more accurately to $\simeq~1\%$.
From the  four parameter fit, we find $R_\mathrm{Drude}\simeq 1.7$~k$\Omega$. Using the expressions 
 \begin{equation}\label{eq:kF_kFl}
 	k_\mathrm{F}=(3\pi^2n)^{1/3} \quad \text{and} \quad \frac{1}{R_\mathrm{Drude}}=d\cdot \frac{e^2}{\hbar}\,\frac{k_\mathrm{F}^2\ell}{3\pi^2}
 \end{equation}
 from the free electron model we estimate $k_\mathrm{F}\ell \simeq 1.9$, a Drude mean free path $\ell\;\simeq0.15$~nm, and an elastic scattering time $\tau\simeq1.3\cdot10^{-16}$~s. Taking the free electron mass, leads to $v_\mathrm{F}\simeq 1.5\cdot10^6$~m/s and a diffusion constant of $D=v_\mathrm{F}\ell/3\simeq0.73$~m$^2$/s, in good agreement with the independent estimate of Eq.~\ref{eq:sup:xi(T)}.

\section{Dirty limit BCS-fit to $J_s(T)$}
\begin{figure}[t]
	\includegraphics[width=.5\textwidth]{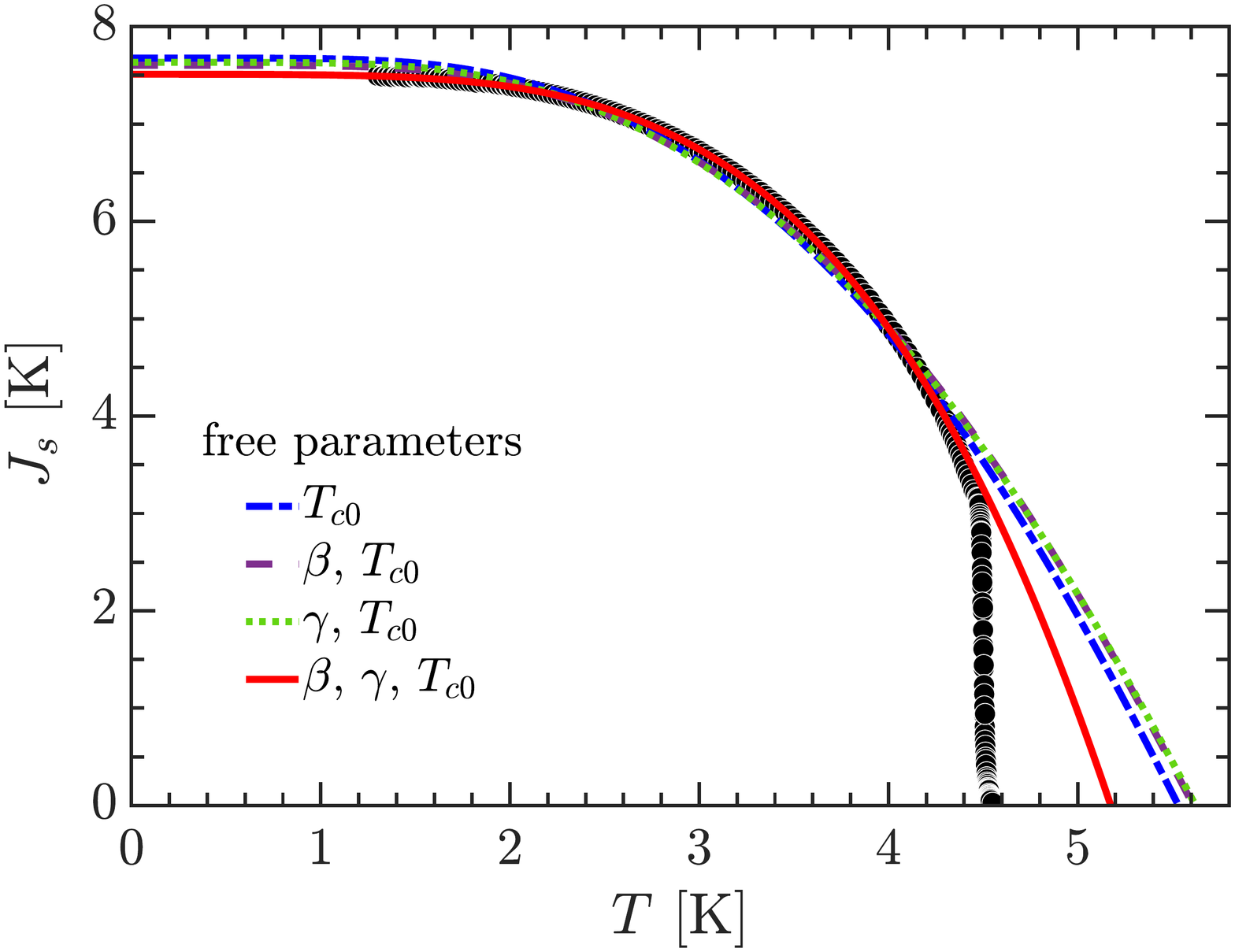}
	\caption[]{Comparison of one, two and three parameter fits of Eq.~\ref{eq:supp:J_BCS} to  $J_s(T)$   in the main text. Three free parameters $T_\mathrm{c0}=5.175$~K, $\beta=0.7304, \gamma= 1.432$ are required to obtain a good agreement between data and fit, see main text.}
	\label{fig:sup:J_s_param_var}
\end{figure}

We fit $J_s(T)$ data using
\begin{align}
	J_s(T)\ &=\  
	J_s(0)\cdot\frac{\Delta(T)}{\Delta(0)}\cdot\tanh\bigg(\frac{\Delta(T)}{2k_\mathrm{B} T}\bigg)\;,
	\label{eq:supp:J_BCS}
\end{align}
%
where $J_s(0) =\beta\cdot J^\mathrm{BCS}_s(0)$. Here, $\beta$ and $\gamma$ are freely adjustable parameters, $J^\mathrm{BCS}_s(0)=\pi\hbar\Delta(T)/(4e^2k_\mathrm{B} R_N)$ is the dirty-limit BCS-expression  \cite{Tinkham1996} for $J_S(0)$, 
$\Delta(0) = \gamma\cdot 1.764\,k_\mathrm{B} T_\mathrm{c0}$,  and $R_N$ being the normal state resistance determined at the maximum in $R(T)$ near 12\,K. As in Refs.~\cite{Yong2013,Mondal_2011b}, we need three free parameters, $T_\mathrm{c0}$, $\beta$, and $\gamma$, to achieve a good fit.  Since $\Delta(0)$ also affects the shape of the curve via argument of $\tanh[\Delta(T)/2k_\mathrm{B} T]$ in Eq.~\ref{eq:supp:J_BCS}, independent variation $\beta$ and $\gamma$ is needed to reproduce the data (purple curve). This becomes evident from fits with one and two free parameters in Fig.~\ref{fig:sup:J_s_param_var}. The one-parameter (red) and two-parameter (blue and green) fits fail to reproduce the curvature of $J_s(T)$ as well as the absolute values of $J_s(0)$ and the independently measured value of $T_\mathrm{c0}$.

\section{High-Field Magnetoresistance}

\begin{figure}[t]
	\includegraphics[width=.5\textwidth]{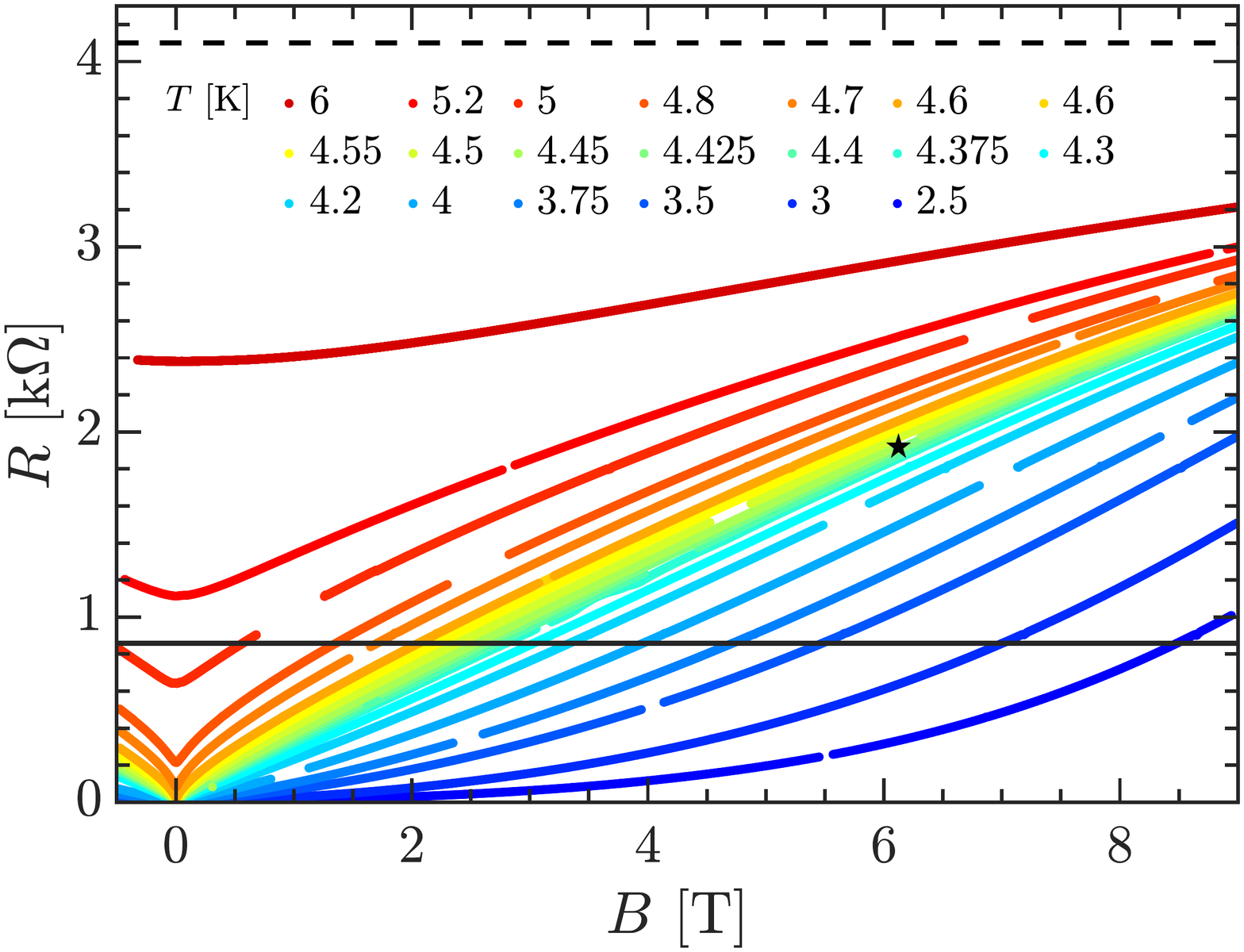}
	\caption[]{Magnetoresistance isotherms. Solid line corresponds to $R(B)=R_t$, see text. Dashed line corresponds to $R=R_N$. Black star denotes $R(B)=0.46R_N$ corresponding to the magnetic field at which scaling is obtained of the  low-field resistance data (Fig.~\ref{fig:R(B)}) in the main text.}
	\label{fig:sup:R(B)}
\end{figure}

Figure~\ref{fig:sup:R(B)} shows magnetoresistance isotherms measured in a  Nb$_3$Sn solenoid  up to 9\,T.  Determination of $B_{c2}(T)$ is difficult because the jump in $R(B)$ is smeared out over a field range of more than 10~T. As criterion for $B_{c2}(T)$, we defined a threshold resistance $R_t=R(B=B_{c2})=0.22R_N$ (solid horizontal line in \ref{fig:sup:R(B)}) such that $B_\mathrm{c2}(T)$ extrapolates to $T_\mathrm{c0}$ (red dots). Note that the value of $B_{c2}(T_\mathrm{BKT})$ extracted from the scaling of the low-field magnetoresistance in Fig.~\ref{fig:R(B)} corresponds to a resistance of $0.46R_N$ and is thus much closer to the standard criterion $0.5R_N$. The latter criterion, however, fails close to $T_\mathrm{c0}$ because the corresponding curve (blue dots) extratolates to much to high temperatures.  From $B_{c2}(T)$ we estimate $\xi_\mathrm{GL}(T)$ via
%
%
%
\begin{align}
	\xi_\mathrm{GL}(T)=\frac{\xi_\mathrm{GL}(0)}{(1-T/T_\mathrm{c0})^{1/2}}=\sqrt{\frac{\hbar D}{\Delta(T)}}\;.
	\label{eq:sup:xi(T)}
\end{align}
with $\xi_\mathrm{GL}(0)=4.26$ nm. Using $\Delta(0)\simeq2.521\cdot  k_\mathrm{B}T_\mathrm{c0}$ we find a diffusion constant of $D\simeq 0.55$~cm$^2$/s  in good agreement with the independent estimate of Eq.~\ref{eq:kF_kFl}.



\begin{figure}[b]
	\includegraphics[width=.45\textwidth]{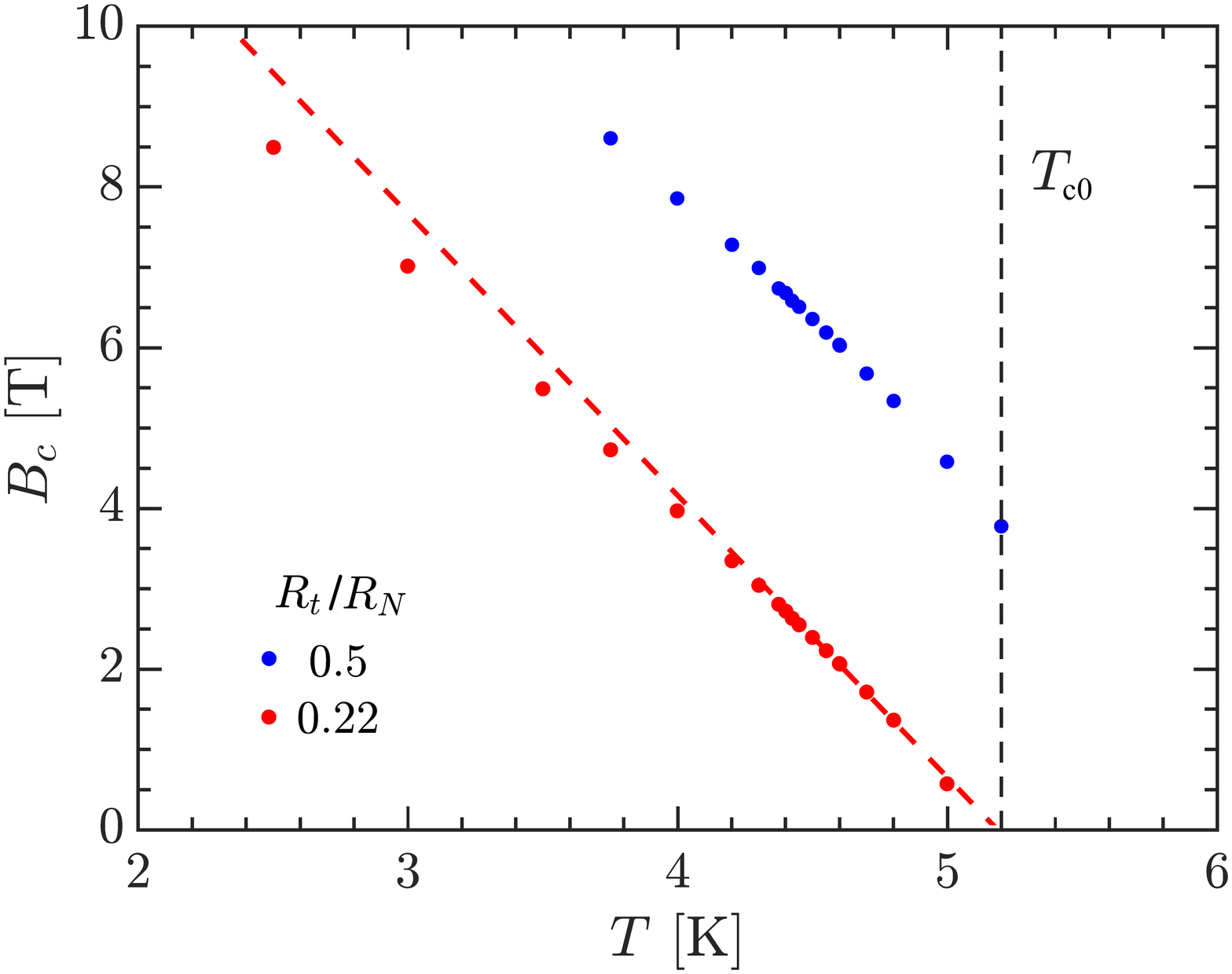}
	\caption[]{Upper critical field $B_{c2}(T)$ according to the criteria  $R_t=0.22R_N$ (red) and $R_t=0.5R_N$ (blue), determined from magnetoresistance isotherms (see text).  The red dashed line is linear fit to $B_{c2}(T)$ close to $T_\mathrm{c0}$. The vertical dashed line indicates $T_\mathrm{c0}=5.175$ K, determined via a BCS fit to $J_s(T)$.}
	\label{fig:sup:bc}
\end{figure}


\begin{figure}[t]
	\includegraphics[width=.48\textwidth]{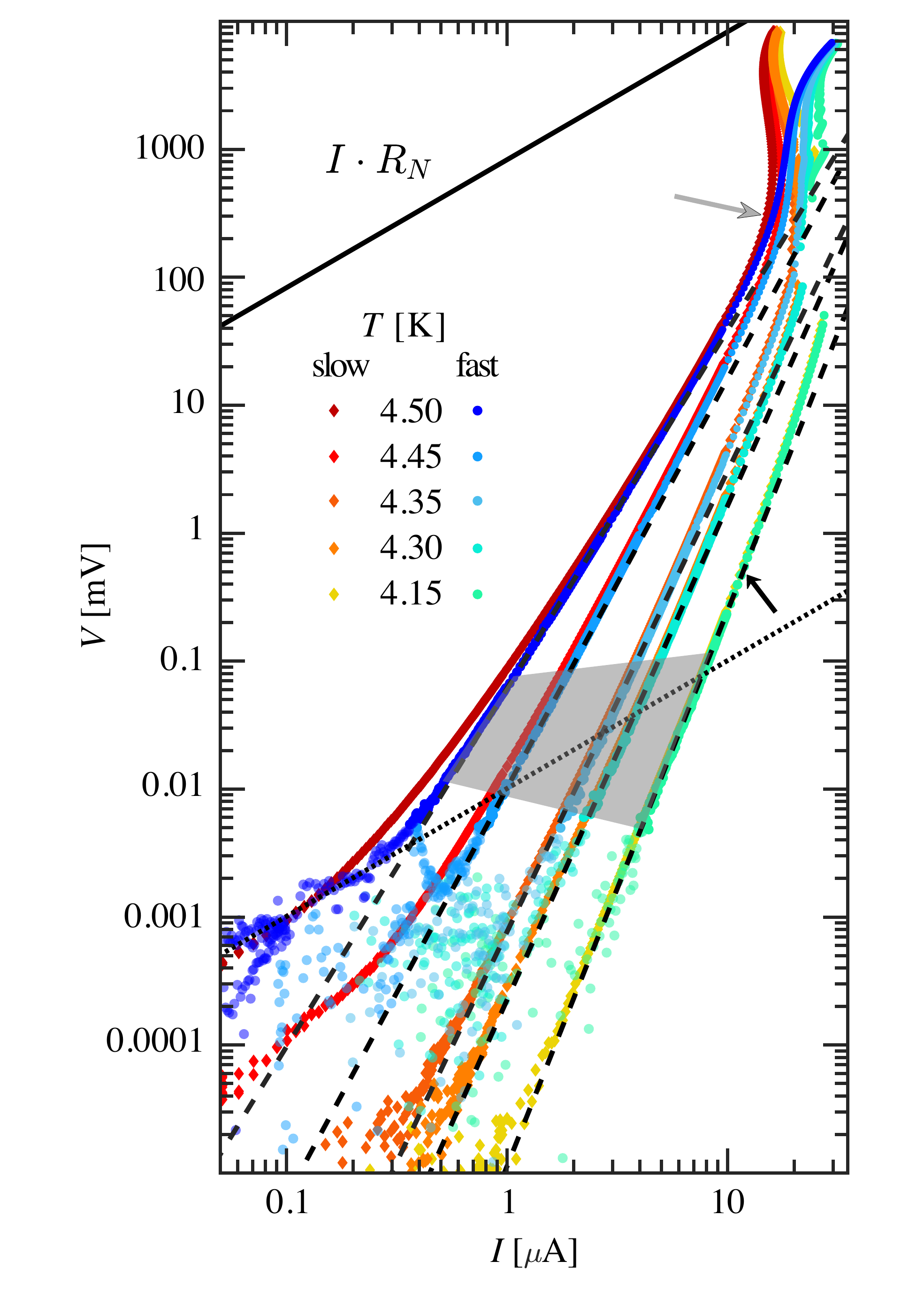}
	\caption[]{$V(I)$ characteristics in the immediate vicinity of $T_\mathrm{BKT}$ measured on the device discussed in the main text. The dot colors indicate both the temperature and the measurement scheme [fast (7 seconds) vs.~slow (10 minutes per sweep)]. Lines correspond to fits using Eq.~\ref{eq:HN-IV} with $\alpha=1$ (dotted) and $\alpha=2.3, 3.2,  3.5, 3.9, 4.3$ (dashed, from  top to bottom). Solid line indicates $V=R_N\cdot I$, where $R_N$ is the normal state resistance. Black arrow points out beginning deviations from HN-behavior due to heating. Grey shaded area displays the fitting range for the extraction of the power law exponent $\alpha(T)$. Grey arrow points at beginning heating of the sample chip in slow sweeps.}
	\label{fig:sup:IV_fast_v_slow}
\end{figure}

\section{$V(I)$-Characteristics}

As discussed in the main text, the superfluid stiffness  $J_s(T)$ can also be extracted from the power-law exponent $\alpha(T)$ of the $V(I)$ characteristics using Halperin-Nelson theory \cite{HalperinNelson_1979} (Figs. \ref{fig:IV} and \ref{fig:L(T)}). Fig.~\ref{fig:sup:IV_fast_v_slow} shows typical $V(I)$ characteristics in a very narrow temperature regime $0.92~T_\mathrm{BKT}<T<1.01~T_\mathrm{BKT}$. In this regime, we employed two different measurement schemes to evaluate the importance of heating effects: Blue curves are fast sweeps of duration 1-7~s, to minimize heating and measurement time. Measurements shown in Fig.~\ref{fig:L(T)} in the main text were performed in the fast scheme. We determine power-law exponents from fast sweeps (red dots in Fig. ~\ref{fig:L(T)}) in the main text (Fig.~\ref{fig:L(T)}) by fitting the data in a region indicated by the shaded grey area to a power law  with exponent~$\alpha(T)$.   Red curves are slow sweeps (duration: ~10 minutes) using a nanovoltmeter. 

We emphasize the importance of extracting power-law exponents at the lowest possible power regime ($I\rightarrow0$, $V\rightarrow0$), to minimize effects of electron heating, which can alter the shape of the $V(I)$-curve. For both the fast and slow measurement scheme, such effects appear  in Fig.~\ref{fig:sup:IV_fast_v_slow} already at power levels of a few picowatts, where data start to deviate from power-law behavior (black arrow), which we attribute to electron overheating \cite{Levinson2019}. At much higher  power $P\sim2~\mu$W (grey arrow) we observe a clear divergence of the slowly measured curves (red) from the fast measured curves (blue). Analysis of data in both electron- and chip-heating regimes will likely result in erroneous values of $\alpha(T)$. Heating effects that push the film towards the normal state can be approximated by power law $V(I)$-characteristics in limited $T$-intervals.  A typical signature of such analysis is a wide temperature range in which $1\leq\alpha\leq3$.  This width, however, should be much smaller than the total width of the fluctuation regime between $T_\mathrm{BKT}$ and the temperatures where $R(T)$ approaches $R_N$. As indicated by the grey shaded area in Fig.~\ref{fig:sup:IV_fast_v_slow}), the  range of $V(I)$ that is governed by the current-induced vortex-anti-vortex depairing is rather small

Above 1\,mV, corresponding to power levels of 10\,pW/square in our devices, $V(I)$-gradually starts to bend upward because of electron heating. In a limited voltage range also these $V(I)$-characteristics can mimick power-law behavior, leading to an apparently broadened transition. At even higher voltage levels $\gtrsim 100\,$mV, or power levels $\gtrsim 2$~\textmu W in the whole meander, heating of the sample stage becomes noticable. This leads to heating instabilities and back-bending of the $V(I)$-characteristics.

Slightly below $T_\mathrm{BKT}$ we observe ohmic tails in the slow measurement at low voltage $V<10^{-6}$ V (red and dark red curve in \ref{fig:sup:IV_fast_v_slow}), that are not discernible in the fast sweeps. As pointed out in \cite{Tamir2019, Benyamini2020}, these tails can result from current noise due to insufficient filtering even at $^4$He-temperatures. In our set-up, we use $\pi$-filters with cutoff frequency 100~MHz for all measurement leads. For the slow measurement rounding towards ohmic behavior occurs at slightly higher voltages when compared to the fast measurement. A clarification of this effect requires further study.

\begin{figure}[t]
	\includegraphics[width=.48\textwidth]{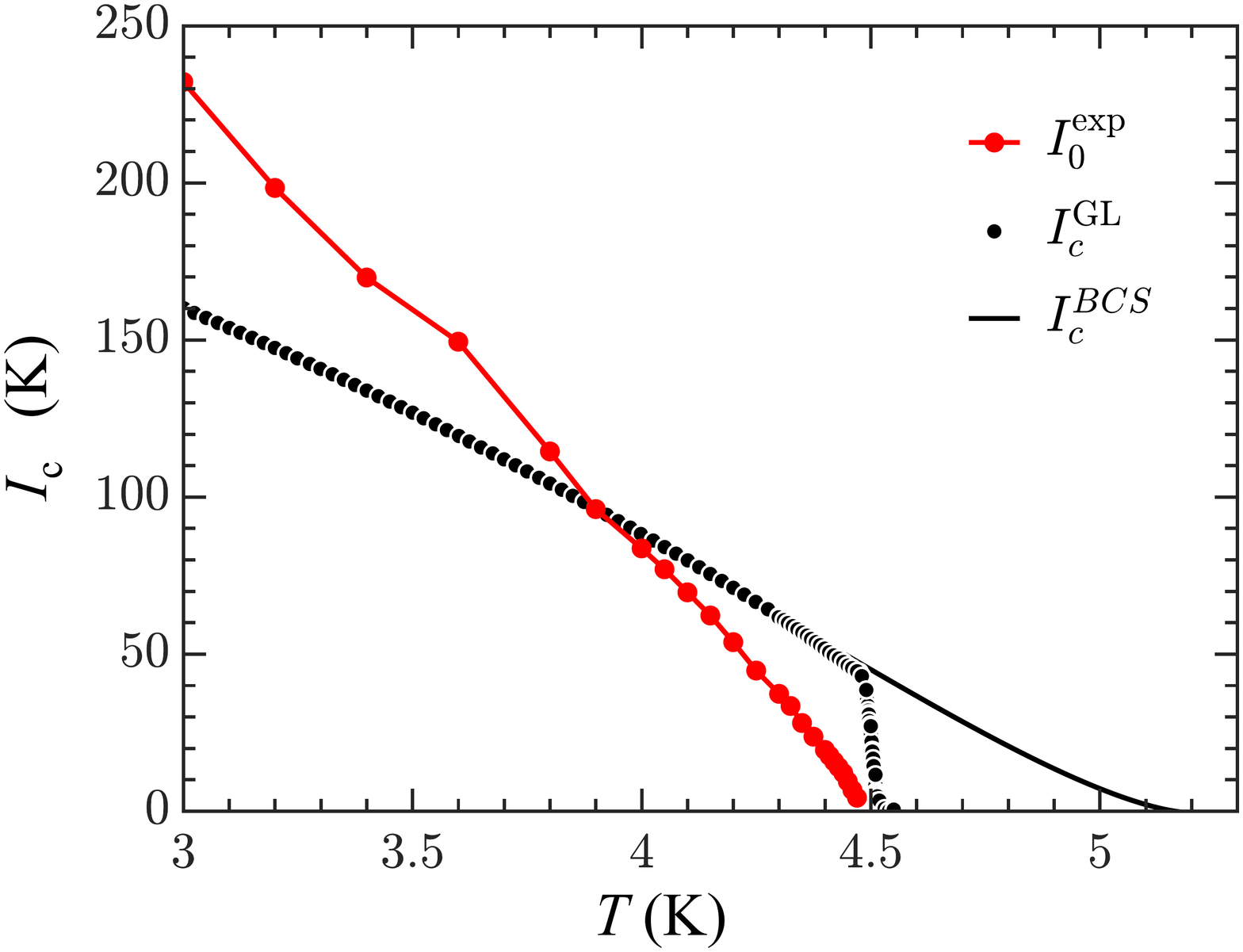}
	\caption[critcur]{Characteristic currents of our films: HN-scaling current $	I_{c}^\mathrm{GL}$ from Eq.~\ref{eq:I_0_IV} (red dots), plotted together with the critical current $I_c^\mathrm{GL}(T)$ according to Eq.~\ref{eq:sup:GL Ic} (black dots), and the standard Ginzburg-Landau critical current $I_c^\mathrm{BCS}$ according to the BCS-extrapolation of $J_s(T)$ towards $T_\mathrm{c0}$ (black line).}
	\label{fig:currents}
\end{figure}

\section{Characteristic Currents}
Another interesting comparison can be performed between the Ginzburg-Landau (GL) critical current $I_{c}^\mathrm{GL}$  and the Halperin-Nelson (HN) scaling current $I_0$ entering the prefactor $A(T)$ in Eq.~\ref{eq:HN-IV}. Withing the error margins of $\xi_\mathrm{GL}$ and $J_s(T)\propto 1/\lambda^2$ we can estimate $I_{c}^\mathrm{GL}$ via
\begin{align}
	I_{c}^\mathrm{GL}(T)\ &=\ \nonumber\frac{\hbar}{3\sqrt{3}e\mu_0}\frac{wd}{\lambda^2(T)\xi(T)}\\[2mm]
	&=\frac{\Phi_0}{3\sqrt{3}\pi}\frac{w}{L_\mathrm{kin}(T)\xi(T)}\;.
	\label{eq:sup:GL Ic}
\end{align}

The error margin is mainly set by the uncertainty of $\xi_\mathrm{GL}$, which is hard to determine reliably from the broad magnetoresistance curves.
The scaling current $I_0^\mathrm{exp}(T)$ is extracted  from the independent measurement of $V(I)$. On the other hand, the full expression for the $V(I)$-characteristics within 
HN theory reads	 \cite{HalperinNelson_1979}:
\begin{align}
	V(I,T)\ =\ I\cdot R_N\cdot\left[2 \pi \frac{J_s(T)}{T}-4\right]\cdot\left[\frac{I}{I_0(T)}\right]^{\alpha(T)-1}\;.
\end{align}
with the exponent  $\alpha(T)-1={\pi J_s(T)}/{T}$.
From this equation we infer the HN scaling current as:
\begin{equation}
	I_0^\mathrm{exp}(T)\ =\ \left[\frac{R_N({2 \alpha(T)-6})}{A(T)}\right]
	\label{eq:I_0_IV}
\end{equation}
where $A(T)$ and $\alpha(T)$ are fit parameters in the fit of double-logarithmic $V(I)$, see also Eq.~\ref{eq:HN-IV} in the main text.

Note  that the definition of $I_0$, given in Ref. \cite{HalperinNelson_1979}
\begin{align}
	I_0^{\mathrm{theo}}(T)\ =\ \frac{wek_\mathrm{B} T_\mathrm{BKT}}{\hbar\xi(T)}
	\label{eq:I_0_theo}
\end{align}
is valid only very near $T_\mathrm{BKT}$.  

In order to slightly generalize Eq.~\ref{eq:I_0_theo} we replace $T_\mathrm{BKT}$ by   $\pi J_s(T\lesssim T_\mathrm{BKT})/2$ and use the relation $J_s(T)\propto1/\lambda^2(T)$ (Eq.~\ref{eq:J_S}), we find that an expression for $I_0(T)$  that reads identical to the GL-critical current (Eq.~\ref{eq:sup:GL Ic}), the only difference being that $1/\lambda^2(T)$ can now be taken form the measurement of $J_s(T)$ rather than assuming the standard GL-form $\lambda(T)=\lambda(0)/\sqrt{1-T/T_\mathrm{c0}}$ in Eq.~\ref{eq:sup:GL Ic}.

When plotting the so-obtained data for $I_c^\mathrm{GL}(T)$ in Fig.~\ref{fig:currents} together with the experimentally determined HN-scaling current $I_0^\mathrm{exp}(T)$ we find a fair agreement. A more accurate derivation of Eq.~\ref{eq:I_0_theo} may provide quantitative understanding also of the prefactor $A(T)$ in Eq.~\ref{eq:HN-IV} in the main text.

\section{Other Devices}
\label{sec:othersamples}
We have performed measurements of $J_s(T)$ on several similar devices made from the same wafer over a period of one year. All devices show a sharp jump of $J_s(T)$ near their intersection with the BKT universal line. They differ in width, while the number of squares was kept close to 100, except for the 10~\textmu m wide device, where if was 200. The variations between the films result from different oxidation states. Over the course of several months the films gradually increase in normal state resistance. Table~1 summarizes the relevant sample parameters.

\begin{table*}[t!]
	\centering
	\renewcommand{\arraystretch}{1.3}
	\setlength{\tabcolsep}{8pt}
	\begin{tabular}{c|ccccccc}
		sample & width [\textmu m] &	$J_s(0) [K]$&	$R_N$ [k$\Omega$] &	$T_{c0}$ [K] &	 $T_\mathrm{BKT}$ [K] &	$\displaystyle\frac{\Delta(0)}{k_BT_\mathrm{c0}}$ &	$\displaystyle\frac{ J_s(0)}{T_\mathrm{c0}}R_N$ [k$\Omega$]\\[2mm]
		\hline\\[-4mm]
		A& 10& 10.726& 3.115&6.167& 5.369& 2.269& 5.418\\
		B&10	&7.981	&3.964&	5.276&	4.601	&2.496	&5.988\\ C&200&	7.700&	$4.167$	&5.183&	4.529&	2.727&	6.190\\
		main text &10&7.511&	4.093&	5.175&	4.488&	2.521&	5.940
	\end{tabular}
	\caption{Sample parameters for several meanders of different width. Sample A was patterned into a freshly made film, while samples B, C, and the sample in the main text come from the same film and were measured several months after deposition in time separations of a few weeks.}
\end{table*}




\begin{thebibliography}{50}%
	\makeatletter
	\providecommand \@ifxundefined [1]{%
		\@ifx{#1\undefined}
	}%
	\providecommand \@ifnum [1]{%
		\ifnum #1\expandafter \@firstoftwo
		\else \expandafter \@secondoftwo
		\fi
	}%
	\providecommand \@ifx [1]{%
		\ifx #1\expandafter \@firstoftwo
		\else \expandafter \@secondoftwo
		\fi
	}%
	\providecommand \natexlab [1]{#1}%
	\providecommand \enquote  [1]{``#1''}%
	\providecommand \bibnamefont  [1]{#1}%
	\providecommand \bibfnamefont [1]{#1}%
	\providecommand \citenamefont [1]{#1}%
	\providecommand \href@noop [0]{\@secondoftwo}%
	\providecommand \href [0]{\begingroup \@sanitize@url \@href}%
	\providecommand \@href[1]{\@@startlink{#1}\@@href}%
	\providecommand \@@href[1]{\endgroup#1\@@endlink}%
	\providecommand \@sanitize@url [0]{\catcode `\\12\catcode `\$12\catcode
		`\&12\catcode `\#12\catcode `\^12\catcode `\_12\catcode `\%12\relax}%
	\providecommand \@@startlink[1]{}%
	\providecommand \@@endlink[0]{}%
	\providecommand \url  [0]{\begingroup\@sanitize@url \@url }%
	\providecommand \@url [1]{\endgroup\@href {#1}{\urlprefix }}%
	\providecommand \urlprefix  [0]{URL }%
	\providecommand \Eprint [0]{\href }%
	\providecommand \doibase [0]{https://doi.org/}%
	\providecommand \selectlanguage [0]{\@gobble}%
	\providecommand \bibinfo  [0]{\@secondoftwo}%
	\providecommand \bibfield  [0]{\@secondoftwo}%
	\providecommand \translation [1]{[#1]}%
	\providecommand \BibitemOpen [0]{}%
	\providecommand \bibitemStop [0]{}%
	\providecommand \bibitemNoStop [0]{.\EOS\space}%
	\providecommand \EOS [0]{\spacefactor3000\relax}%
	\providecommand \BibitemShut  [1]{\csname bibitem#1\endcsname}%
	\let\auto@bib@innerbib\@empty
	\bibitem [{\citenamefont {Kosterlitz}\ and\ \citenamefont
		{Thouless}(1973)}]{KT_1973}%
	\BibitemOpen
	\bibfield  {author} {\bibinfo {author} {\bibfnamefont {J.~M.}\ \bibnamefont
			{Kosterlitz}}\ and\ \bibinfo {author} {\bibfnamefont {D.~J.}\ \bibnamefont
			{Thouless}},\ }\bibfield  {title} {\bibinfo {title} {Ordering, metastability
			and phase transitions in two-dimensional systems},\ }\href@noop {} {\bibfield
		{journal} {\bibinfo  {journal} {J. Phys. C: Solid State Phys.}\ }\textbf
		{\bibinfo {volume} {6}},\ \bibinfo {pages} {1181} (\bibinfo {year}
		{1973})}\BibitemShut {NoStop}%
	\bibitem [{\citenamefont {Kosterlitz}(1974)}]{Kosterlitz_1974}%
	\BibitemOpen
	\bibfield  {author} {\bibinfo {author} {\bibfnamefont {J.~M.}\ \bibnamefont
			{Kosterlitz}},\ }\bibfield  {title} {\bibinfo {title} {The critical
			properties of the two-dimensional XY-model},\ }\href@noop {} {\bibfield
		{journal} {\bibinfo  {journal} {J. Phys. C: Solid State Phys.}\ }\textbf
		{\bibinfo {volume} {7}},\ \bibinfo {pages} {1046} (\bibinfo {year}
		{1974})}\BibitemShut {NoStop}%
	\bibitem [{\citenamefont {Nelson}\ and\ \citenamefont
		{Kosterlitz}(1977)}]{NelsonKosterlitz1977}%
	\BibitemOpen
	\bibfield  {author} {\bibinfo {author} {\bibfnamefont {D.~R.}\ \bibnamefont
			{Nelson}}\ and\ \bibinfo {author} {\bibfnamefont {J.~M.}\ \bibnamefont
			{Kosterlitz}},\ }\bibfield  {title} {\bibinfo {title} {{Universal Jump in the
				Superfluid Density of Two-Dimensional Superfluids}},\ }\href
	{https://doi.org/10.1103/PhysRevLett.39.1201} {\bibfield  {journal} {\bibinfo
			{journal} {Phys. Rev. Lett.}\ }\textbf {\bibinfo {volume} {39}},\ \bibinfo
		{pages} {1201} (\bibinfo {year} {1977})}\BibitemShut {NoStop}%
	\bibitem [{\citenamefont {McQueeney}\ \emph {et~al.}(1984)\citenamefont
		{McQueeney}, \citenamefont {Agnolet},\ and\ \citenamefont
		{Reppy}}]{McQueeney1984}%
	\BibitemOpen
	\bibfield  {author} {\bibinfo {author} {\bibfnamefont {D.}~\bibnamefont
			{McQueeney}}, \bibinfo {author} {\bibfnamefont {G.}~\bibnamefont {Agnolet}},\
		and\ \bibinfo {author} {\bibfnamefont {J.~D.}\ \bibnamefont {Reppy}},\
	}\bibfield  {title} {\bibinfo {title} {{Surface Superfluidity in Dilute
				$^4$He-$^3$He Mixtures}},\ }\href@noop {} {\bibfield  {journal} {\bibinfo
			{journal} {Phys. Rev. Lett.}\ }\textbf {\bibinfo {volume} {52}},\ \bibinfo
		{pages} {1325} (\bibinfo {year} {1984})}\BibitemShut {NoStop}%
	%
	\bibitem{AHNS1} V.~Ambegaokar, B.\,I.~Halperin, D.\,R.~Nelson, E.\,D.~Siggia, Dissipation in Two-Dimensional Superfluids, Phys.~Rev.~Lett.~{\bf 40}, 783 (1978);
	%
	\bibitem{AHNS2} V.~Ambegaokar, B.~I.~Halperin, D.~R.~Nelson, E.~D.~Siggia, 
	Dynamics of superfluid films, Phys.~Rev.~B {\bf 21}, 1806 (1980).
	%
	\bibitem [{\citenamefont {Halperin}\ and\ \citenamefont
		{Nelson}(1979)}]{HalperinNelson_1979}%
	\BibitemOpen
	\bibfield  {author} {\bibinfo {author} {\bibfnamefont {B.~I.}\ \bibnamefont
			{Halperin}}\ and\ \bibinfo {author} {\bibfnamefont {D.~R.}\ \bibnamefont
			{Nelson}},\ }\bibfield  {title} {\bibinfo {title} {{Resistive Transition in
				Superconducting Films}},\ }\href@noop {} {\bibfield  {journal} {\bibinfo
			{journal} {{J.~Low Temp.~Phys.}}\ }\textbf {\bibinfo {volume} {36}},\
		\bibinfo {pages} {599} (\bibinfo {year} {1979})}\BibitemShut {NoStop}%
	%
	\bibitem [{\citenamefont {Benfatto}\ \emph {et~al.}(2009)\citenamefont
		{Benfatto}, \citenamefont {Castellani},\ and\ \citenamefont
		{Giamarchi}}]{Benfatto2009}%
	\BibitemOpen
	\bibfield  {author} {\bibinfo {author} {\bibfnamefont {L.}~\bibnamefont
			{Benfatto}}, \bibinfo {author} {\bibfnamefont {C.}~\bibnamefont
			{Castellani}},\ and\ \bibinfo {author} {\bibfnamefont {T.}~\bibnamefont
			{Giamarchi}},\ }\bibfield  {title} {\bibinfo {title} {{Broadening of the
				Berezinskii-Kosterlitz-Thouless superconducting transition by inhomogeneity
				and finite-size effects}},\ }\href
	{https://doi.org/10.1103/PhysRevB.80.214506} {\bibfield  {journal} {\bibinfo
			{journal} {Phys. Rev. B}\ }\textbf {\bibinfo {volume} {80}},\ \bibinfo
		{pages} {214506} (\bibinfo {year} {2009})}\BibitemShut {NoStop}%
	\bibitem [{\citenamefont {A.~T.~Fiory}\ and\ \citenamefont
		{Glaberson}(1983)}]{FioryHebardGlaberson_1983}%
	\BibitemOpen
	\bibfield  {author} {\bibinfo {author}
		\bibnamefont {A.~T.~Fiory}},\  \bibinfo {author} \ {\bibfnamefont {A.~F.~Hebard}\ and\ \bibinfo {author} {\bibfnamefont {W.~I.}\
			\bibnamefont {Glaberson}},\ }\bibfield  {title} {\bibinfo {title}
		{Superconducting phase transitions in indium/indium-oxide thin-film
			composites},\ }\href@noop {} {\bibfield  {journal} {\bibinfo  {journal}
			{Phys. Rev. B}\ }\textbf {\bibinfo {volume} {28}},\ \bibinfo {pages} {5075}
		(\bibinfo {year} {1983})}\BibitemShut {NoStop}%
	\bibitem [{\citenamefont {Yong}\ \emph {et~al.}(2013)\citenamefont {Yong},
		\citenamefont {Lemberger}, \citenamefont {Benfatto}, \citenamefont {Ilin},\
		and\ \citenamefont {Siegel}}]{Yong2013}%
	\BibitemOpen
	\bibfield  {author} {\bibinfo {author} {\bibfnamefont {J.}~\bibnamefont
			{Yong}}, \bibinfo {author} {\bibfnamefont {T.~R.}\ \bibnamefont {Lemberger}},
		\bibinfo {author} {\bibfnamefont {L.}~\bibnamefont {Benfatto}}, \bibinfo
		{author} {\bibfnamefont {K.}~\bibnamefont {Ilin}},\ and\ \bibinfo {author}
		{\bibfnamefont {M.}~\bibnamefont {Siegel}},\ }\bibfield  {title} {\bibinfo
		{title} {{Robustness of the Berezinskii-Kosterlitz-Thouless transition in
				ultrathin NbN films near the superconductor-insulator transition}},\ }\href
	{https://doi.org/10.1103/PhysRevB.87.184505} {\bibfield  {journal} {\bibinfo
			{journal} {Phys. Rev. B}\ }\textbf {\bibinfo {volume} {87}},\ \bibinfo
		{pages} {184505} (\bibinfo {year} {2013})}\BibitemShut {NoStop}%
	\bibitem [{\citenamefont {Venditti}\ \emph {et~al.}(2019)\citenamefont
		{Venditti}, \citenamefont {Biscaras}, \citenamefont {Hurand}, \citenamefont
		{Bergeal}, \citenamefont {Lesueur}, \citenamefont {Dogra}, \citenamefont
		{Budhani}, \citenamefont {Mondal}, \citenamefont {Jesudasan}, \citenamefont
		{Raychaudhuri}, \citenamefont {Caprara},\ and\ \citenamefont
		{Benfatto}}]{Venditti2019}%
	\BibitemOpen
	\bibfield  {author} {\bibinfo {author} {\bibfnamefont {G.}~\bibnamefont
			{Venditti}}, \bibinfo {author} {\bibfnamefont {J.}~\bibnamefont {Biscaras}},
		\bibinfo {author} {\bibfnamefont {S.}~\bibnamefont {Hurand}}, \bibinfo
		{author} {\bibfnamefont {N.}~\bibnamefont {Bergeal}}, \bibinfo {author}
		{\bibfnamefont {J.}~\bibnamefont {Lesueur}}, \bibinfo {author} {\bibfnamefont
			{A.}~\bibnamefont {Dogra}}, \bibinfo {author} {\bibfnamefont {R.~C.}\
			\bibnamefont {Budhani}}, \bibinfo {author} {\bibfnamefont {M.}~\bibnamefont
			{Mondal}}, \bibinfo {author} {\bibfnamefont {J.}~\bibnamefont {Jesudasan}},
		\bibinfo {author} {\bibfnamefont {P.}~\bibnamefont {Raychaudhuri}}, \bibinfo
		{author} {\bibfnamefont {S.}~\bibnamefont {Caprara}},\ and\ \bibinfo {author}
		{\bibfnamefont {L.}~\bibnamefont {Benfatto}},\ }\bibfield  {title} {\bibinfo
		{title} {{Nonlinear $I\text{\ensuremath{-}}V$ characteristics of
				two-dimensional superconductors: Berezinskii-Kosterlitz-Thouless physics
				versus inhomogeneity}},\ }\href {https://doi.org/10.1103/PhysRevB.100.064506}
	{\bibfield  {journal} {\bibinfo  {journal} {Phys. Rev. B}\ }\textbf {\bibinfo
			{volume} {100}},\ \bibinfo {pages} {064506} (\bibinfo {year}
		{2019})}\BibitemShut {NoStop}%
	\bibitem [{\citenamefont {Turneaure}\ \emph {et~al.}(2000)\citenamefont
		{Turneaure}, \citenamefont {Lemberger},\ and\ \citenamefont
		{Graybeal}}]{Turneaure2000}%
	\BibitemOpen
	\bibfield  {author} {\bibinfo {author} {\bibfnamefont {S.~J.}\ \bibnamefont
			{Turneaure}}, \bibinfo {author} {\bibfnamefont {T.~R.}\ \bibnamefont
			{Lemberger}},\ and\ \bibinfo {author} {\bibfnamefont {J.~M.}\ \bibnamefont
			{Graybeal}},\ }\bibfield  {title} {\bibinfo {title} {{Effect of Thermal Phase
				Fluctuations on the Superfluid Density of Two-Dimensional Superconducting
				Films}},\ }\href {https://doi.org/10.1103/PhysRevLett.84.987} {\bibfield
		{journal} {\bibinfo  {journal} {Phys. Rev. Lett.}\ }\textbf {\bibinfo
			{volume} {84}},\ \bibinfo {pages} {987} (\bibinfo {year} {2000})}\BibitemShut
	{NoStop}%
	\bibitem [{\citenamefont {Mondal}\ \emph
		{et~al.}(2011{\natexlab{a}})\citenamefont {Mondal}, \citenamefont {Kumar},
		\citenamefont {Chand}, \citenamefont {Kamlapure}, \citenamefont {Saraswat},
		\citenamefont {Seibold}, \citenamefont {Benfatto},\ and\ \citenamefont
		{Raychaudhuri}}]{Mondal_2011b}%
	\BibitemOpen
	\bibfield  {author} {\bibinfo {author} {\bibfnamefont {M.}~\bibnamefont
			{Mondal}}, \bibinfo {author} {\bibfnamefont {S.}~\bibnamefont {Kumar}},
		\bibinfo {author} {\bibfnamefont {M.}~\bibnamefont {Chand}}, \bibinfo
		{author} {\bibfnamefont {A.}~\bibnamefont {Kamlapure}}, \bibinfo {author}
		{\bibfnamefont {G.}~\bibnamefont {Saraswat}}, \bibinfo {author}
		{\bibfnamefont {G.}~\bibnamefont {Seibold}}, \bibinfo {author} {\bibfnamefont
			{L.}~\bibnamefont {Benfatto}},\ and\ \bibinfo {author} {\bibfnamefont
			{P.}~\bibnamefont {Raychaudhuri}},\ }\bibfield  {title} {\bibinfo {title}
		{{Role of the Vortex-Core Energy on the Berezinskii-Kosterlitz-Thouless
				Transition in Thin Films of NbN}},\ }\href
	{https://doi.org/10.1103/PhysRevLett.107.217003} {\bibfield  {journal}
		{\bibinfo  {journal} {Phys. Rev. Lett.}\ }\textbf {\bibinfo {volume} {107}},\
		\bibinfo {pages} {217003} (\bibinfo {year} {2011}{\natexlab{a}})}\BibitemShut
	{NoStop}%
	\bibitem [{\citenamefont {Ghosal}\ \emph
		{et~al.}(1998{\natexlab{a}})\citenamefont {Ghosal}, \citenamefont
		{Randeria},\ and\ \citenamefont {Trivedi}}]{Ghosal1998}%
	\BibitemOpen
	\bibfield  {author} {\bibinfo {author} {\bibfnamefont {A.}~\bibnamefont
			{Ghosal}}, \bibinfo {author} {\bibfnamefont {M.}~\bibnamefont {Randeria}},\
		and\ \bibinfo {author} {\bibfnamefont {N.}~\bibnamefont {Trivedi}},\
	}\bibfield  {title} {\bibinfo {title} {Role of spatial amplitude fluctuations
			in highly disordered s-wave superconductors},\ }\href@noop {} {\bibfield
		{journal} {\bibinfo  {journal} {Phys. Rev. Lett.}\ }\textbf {\bibinfo
			{volume} {81}},\ \bibinfo {pages} {3940} (\bibinfo {year}
		{1998}{\natexlab{a}})}\BibitemShut {NoStop}%
	\bibitem [{\citenamefont {Ghosal}\ \emph
		{et~al.}(2001{\natexlab{b}})\citenamefont {Ghosal}, \citenamefont
		{Randeria},\ and\ \citenamefont {Trivedi}}]{Ghosal2001b}%
	\BibitemOpen
	\bibfield  {author} {\bibinfo {author} {\bibfnamefont {A.}~\bibnamefont
			{Ghosal}}, \bibinfo {author} {\bibfnamefont {M.}~\bibnamefont {Randeria}},\
		and\ \bibinfo {author} {\bibfnamefont {N.}~\bibnamefont {Trivedi}},\
	}\bibfield  {title} {\bibinfo {title} {Inhomogeneous pairing in highly
			disordered s-wave superconductors},\ }\href@noop {} {\bibfield  {journal}
		{\bibinfo  {journal} {Phys. Rev. B}\ }\textbf {\bibinfo {volume} {65}},\
		\bibinfo {pages} {014501} (\bibinfo {year} {2001}{\natexlab{b}})}\BibitemShut
	{NoStop}%
	\bibitem [{\citenamefont {Sac\'ep\'e}\ \emph {et~al.}(2008)\citenamefont
		{Sac\'ep\'e}, \citenamefont {Chapelier}, \citenamefont {Baturina},
		\citenamefont {Vinokur}, \citenamefont {Baklanov},\ and\ \citenamefont
		{Sanquer}}]{Sacepe2008}%
	\BibitemOpen
	\bibfield  {author} {\bibinfo {author} {\bibfnamefont {B.}~\bibnamefont
			{Sac\'ep\'e}}, \bibinfo {author} {\bibfnamefont {C.}~\bibnamefont
			{Chapelier}}, \bibinfo {author} {\bibfnamefont {T.~I.}\ \bibnamefont
			{Baturina}}, \bibinfo {author} {\bibfnamefont {V.~M.}\ \bibnamefont
			{Vinokur}}, \bibinfo {author} {\bibfnamefont {M.~R.}\ \bibnamefont
			{Baklanov}},\ and\ \bibinfo {author} {\bibfnamefont {M.}~\bibnamefont
			{Sanquer}},\ }\bibfield  {title} {\bibinfo {title} {Disorder-induced
			inhomogeneities of the superconducting state close to the
			superconductor-insulator transition},\ }\href
	{https://doi.org/10.1103/PhysRevLett.101.157006} {\bibfield  {journal}
		{\bibinfo  {journal} {Phys. Rev. Lett.}\ }\textbf {\bibinfo {volume} {101}},\
		\bibinfo {pages} {157006} (\bibinfo {year} {2008})}\BibitemShut {NoStop}%
	\bibitem [{\citenamefont {Carbillet}\ \emph {et~al.}(2016)\citenamefont
		{Carbillet}, \citenamefont {Caprara}, \citenamefont {Grilli}, \citenamefont
		{Brun}, \citenamefont {Cren}, \citenamefont {Debontridder}, \citenamefont
		{Vignolle}, \citenamefont {Tabis}, \citenamefont {Demaille}, \citenamefont
		{Largeau}, \citenamefont {Ilin}, \citenamefont {Siegel}, \citenamefont
		{Roditchev},\ and\ \citenamefont {Leridon}}]{Carbillet2016}%
	\BibitemOpen
	\bibfield  {author} {\bibinfo {author} {\bibfnamefont {C.}~\bibnamefont
			{Carbillet}}, \bibinfo {author} {\bibfnamefont {S.}~\bibnamefont {Caprara}},
		\bibinfo {author} {\bibfnamefont {M.}~\bibnamefont {Grilli}}, \bibinfo
		{author} {\bibfnamefont {C.}~\bibnamefont {Brun}}, \bibinfo {author}
		{\bibfnamefont {T.}~\bibnamefont {Cren}}, \bibinfo {author} {\bibfnamefont
			{F.}~\bibnamefont {Debontridder}}, \bibinfo {author} {\bibfnamefont
			{B.}~\bibnamefont {Vignolle}}, \bibinfo {author} {\bibfnamefont
			{W.}~\bibnamefont {Tabis}}, \bibinfo {author} {\bibfnamefont
			{D.}~\bibnamefont {Demaille}}, \bibinfo {author} {\bibfnamefont
			{L.}~\bibnamefont {Largeau}}, \bibinfo {author} {\bibfnamefont
			{K.}~\bibnamefont {Ilin}}, \bibinfo {author} {\bibfnamefont {M.}~\bibnamefont
			{Siegel}}, \bibinfo {author} {\bibfnamefont {D.}~\bibnamefont {Roditchev}},\
		and\ \bibinfo {author} {\bibfnamefont {B.}~\bibnamefont {Leridon}},\
	}\bibfield  {title} {\bibinfo {title} {Confinement of superconducting
			fluctuations due to emergent electronic inhomogeneities},\ }\href
	{https://doi.org/10.1103/PhysRevB.93.144509} {\bibfield  {journal} {\bibinfo
			{journal} {Phys. Rev. B}\ }\textbf {\bibinfo {volume} {93}},\ \bibinfo
		{pages} {144509} (\bibinfo {year} {2016})}\BibitemShut {NoStop}%
	\bibitem [{\citenamefont {Carbillet}\ \emph {et~al.}(2020)\citenamefont
		{Carbillet}, \citenamefont {Cherkez}, \citenamefont {Skvortsov},
		\citenamefont {Feigel'man}, \citenamefont {Debontridder}, \citenamefont
		{Ioffe}, \citenamefont {Stolyarov}, \citenamefont {Ilin}, \citenamefont
		{Siegel}, \citenamefont {Roditchev}, \citenamefont {Cren},\ and\
		\citenamefont {Brun}}]{Carbillet2020}%
	\BibitemOpen
	\bibfield  {author} {\bibinfo {author} {\bibfnamefont {C.}~\bibnamefont
			{Carbillet}}, \bibinfo {author} {\bibfnamefont {V.}~\bibnamefont {Cherkez}},
		\bibinfo {author} {\bibfnamefont {M.~A.}\ \bibnamefont {Skvortsov}}, \bibinfo
		{author} {\bibfnamefont {M.~V.}\ \bibnamefont {Feigel'man}}, \bibinfo
		{author} {\bibfnamefont {F.}~\bibnamefont {Debontridder}}, \bibinfo {author}
		{\bibfnamefont {L.~B.}\ \bibnamefont {Ioffe}}, \bibinfo {author}
		{\bibfnamefont {V.~S.}\ \bibnamefont {Stolyarov}}, \bibinfo {author}
		{\bibfnamefont {K.}~\bibnamefont {Ilin}}, \bibinfo {author} {\bibfnamefont
			{M.}~\bibnamefont {Siegel}}, \bibinfo {author} {\bibfnamefont
			{D.}~\bibnamefont {Roditchev}}, \bibinfo {author} {\bibfnamefont
			{T.}~\bibnamefont {Cren}},\ and\ \bibinfo {author} {\bibfnamefont
			{C.}~\bibnamefont {Brun}},\ }\bibfield  {title} {\bibinfo {title}
		{Spectroscopic evidence for strong correlations between local superconducting
			gap and local altshuler-aronov density of states suppression in ultrathin nbn
			films},\ }\href {https://doi.org/10.1103/PhysRevB.102.024504} {\bibfield
		{journal} {\bibinfo  {journal} {Phys. Rev. B}\ }\textbf {\bibinfo {volume}
			{102}},\ \bibinfo {pages} {024504} (\bibinfo {year} {2020})}\BibitemShut
	{NoStop}%
	\bibitem [{\citenamefont {Stosiek}\ \emph {et~al.}(2020)\citenamefont
		{Stosiek}, \citenamefont {Lang},\ and\ \citenamefont {Evers}}]{Stosiek2020}%
	\BibitemOpen
	\bibfield  {author} {\bibinfo {author} {\bibfnamefont {M.}~\bibnamefont
			{Stosiek}}, \bibinfo {author} {\bibfnamefont {B.}~\bibnamefont {Lang}},\ and\
		\bibinfo {author} {\bibfnamefont {F.}~\bibnamefont {Evers}},\ }\bibfield
	{title} {\bibinfo {title} {Self-consistent-field ensembles of disordered
			hamiltonians: Efficient solver and application to superconducting films},\
	}\href {https://doi.org/10.1103/PhysRevB.101.144503} {\bibfield  {journal}
		{\bibinfo  {journal} {Phys. Rev. B}\ }\textbf {\bibinfo {volume} {101}},\
		\bibinfo {pages} {144503} (\bibinfo {year} {2020})}\BibitemShut {NoStop}%
	\bibitem [{\citenamefont {Crane}\ \emph {et~al.}(2007)\citenamefont {Crane},
		\citenamefont {Armitage}, \citenamefont {Johansson}, \citenamefont
		{Sambandamurthy}, \citenamefont {Shahar},\ and\ \citenamefont
		{Gr\"uner}}]{Crane2007}%
	\BibitemOpen
	\bibfield  {author} {\bibinfo {author} {\bibfnamefont {R.~W.}\ \bibnamefont
			{Crane}}, \bibinfo {author} {\bibfnamefont {N.~P.}\ \bibnamefont {Armitage}},
		\bibinfo {author} {\bibfnamefont {A.}~\bibnamefont {Johansson}}, \bibinfo
		{author} {\bibfnamefont {G.}~\bibnamefont {Sambandamurthy}}, \bibinfo
		{author} {\bibfnamefont {D.}~\bibnamefont {Shahar}},\ and\ \bibinfo {author}
		{\bibfnamefont {G.}~\bibnamefont {Gr\"uner}},\ }\bibfield  {title} {\bibinfo
		{title} {Fluctuations, dissipation, and nonuniversal superfluid jumps in
			two-dimensional superconductors},\ }\href
	{https://doi.org/10.1103/PhysRevB.75.094506} {\bibfield  {journal} {\bibinfo
			{journal} {Phys. Rev. B}\ }\textbf {\bibinfo {volume} {75}},\ \bibinfo
		{pages} {094506(R)} (\bibinfo {year} {2007})}\BibitemShut {NoStop}%
	\bibitem [{\citenamefont {Mandal}\ \emph {et~al.}(2020)\citenamefont {Mandal},
		\citenamefont {Dutta}, \citenamefont {Basistha}, \citenamefont {Roy},
		\citenamefont {Jesudasan}, \citenamefont {Bagwe}, \citenamefont {Benfatto},
		\citenamefont {Thamizhavel},\ and\ \citenamefont
		{Raychaudhuri}}]{Mandal2020}%
	\BibitemOpen
	\bibfield  {author} {\bibinfo {author} {\bibfnamefont {S.}~\bibnamefont
			{Mandal}}, \bibinfo {author} {\bibfnamefont {S.}~\bibnamefont {Dutta}},
		\bibinfo {author} {\bibfnamefont {S.}~\bibnamefont {Basistha}}, \bibinfo
		{author} {\bibfnamefont {I.}~\bibnamefont {Roy}}, \bibinfo {author}
		{\bibfnamefont {J.}~\bibnamefont {Jesudasan}}, \bibinfo {author}
		{\bibfnamefont {V.}~\bibnamefont {Bagwe}}, \bibinfo {author} {\bibfnamefont
			{L.}~\bibnamefont {Benfatto}}, \bibinfo {author} {\bibfnamefont
			{A.}~\bibnamefont {Thamizhavel}},\ and\ \bibinfo {author} {\bibfnamefont
			{P.}~\bibnamefont {Raychaudhuri}},\ }\bibfield  {title} {\bibinfo {title}
		{{Destruction of superconductivity through phase fluctuations in ultrathin
				$a$-MoGe films}},\ }\href {https://doi.org/10.1103/PhysRevB.102.060501}
	{\bibfield  {journal} {\bibinfo  {journal} {Phys. Rev. B}\ }\textbf {\bibinfo
			{volume} {102}},\ \bibinfo {pages} {060501(R)} (\bibinfo {year}
		{2020})}\BibitemShut {NoStop}%
	\bibitem [{\citenamefont {Broun}\ \emph {et~al.}(2007)\citenamefont {Broun},
		\citenamefont {Huttema}, \citenamefont {Turner}, \citenamefont {\"Ozcan},
		\citenamefont {Morgan}, \citenamefont {Liang}, \citenamefont {Hardy},\ and\
		\citenamefont {Bonn}}]{Broun2007}%
	\BibitemOpen
	\bibfield  {author} {\bibinfo {author} {\bibfnamefont {D.~M.}\ \bibnamefont
			{Broun}}, \bibinfo {author} {\bibfnamefont {W.~A.}\ \bibnamefont {Huttema}},
		\bibinfo {author} {\bibfnamefont {P.~J.}\ \bibnamefont {Turner}}, \bibinfo
		{author} {\bibfnamefont {S.}~\bibnamefont {\"Ozcan}}, \bibinfo {author}
		{\bibfnamefont {B.}~\bibnamefont {Morgan}}, \bibinfo {author} {\bibfnamefont
			{R.}~\bibnamefont {Liang}}, \bibinfo {author} {\bibfnamefont {W.~N.}\
			\bibnamefont {Hardy}},\ and\ \bibinfo {author} {\bibfnamefont {D.~A.}\
			\bibnamefont {Bonn}},\ }\bibfield  {title} {\bibinfo {title} {{Superfluid
				Density in a Highly Underdoped
				${\mathrm{YBa}}_{2}{\mathrm{Cu}}_{3}{\mathrm{O}}_{6+y}$ Superconductor}},\
	}\href {https://doi.org/10.1103/PhysRevLett.99.237003} {\bibfield  {journal}
		{\bibinfo  {journal} {Phys. Rev. Lett.}\ }\textbf {\bibinfo {volume} {99}},\
		\bibinfo {pages} {237003} (\bibinfo {year} {2007})}\BibitemShut {NoStop}%
	\bibitem [{\citenamefont {Kamal}\ \emph {et~al.}(1994)\citenamefont {Kamal},
		\citenamefont {Bonn}, \citenamefont {Goldenfeld}, \citenamefont {Hirschfeld},
		\citenamefont {Liang},\ and\ \citenamefont {Hardy}}]{Kamal1994}%
	\BibitemOpen
	\bibfield  {author} {\bibinfo {author} {\bibfnamefont {S.}~\bibnamefont
			{Kamal}}, \bibinfo {author} {\bibfnamefont {D.~A.}\ \bibnamefont {Bonn}},
		\bibinfo {author} {\bibfnamefont {N.}~\bibnamefont {Goldenfeld}}, \bibinfo
		{author} {\bibfnamefont {P.~J.}\ \bibnamefont {Hirschfeld}}, \bibinfo
		{author} {\bibfnamefont {R.}~\bibnamefont {Liang}},\ and\ \bibinfo {author}
		{\bibfnamefont {W.~N.}\ \bibnamefont {Hardy}},\ }\bibfield  {title} {\bibinfo
		{title} {{Penetration Depth Measurements of 3D $\mathrm{XY}$ Critical
				Behavior in ${\mathrm{YBa}}_{2}$${\mathrm{Cu}}_{3}$${\mathrm{O}}_{6.95}$
				Crystals}},\ }\href {https://doi.org/10.1103/PhysRevLett.73.1845} {\bibfield
		{journal} {\bibinfo  {journal} {Phys. Rev. Lett.}\ }\textbf {\bibinfo
			{volume} {73}},\ \bibinfo {pages} {1845} (\bibinfo {year}
		{1994})}\BibitemShut {NoStop}%
	\bibitem [{\citenamefont {Yong}\ \emph {et~al.}(2012)\citenamefont {Yong},
		\citenamefont {Hinton}, \citenamefont {McCray}, \citenamefont {Randeria},
		\citenamefont {Naamneh}, \citenamefont {Kanigel},\ and\ \citenamefont
		{Lemberger}}]{Yong2012}%
	\BibitemOpen
	\bibfield  {author} {\bibinfo {author} {\bibfnamefont {J.}~\bibnamefont
			{Yong}}, \bibinfo {author} {\bibfnamefont {M.~J.}\ \bibnamefont {Hinton}},
		\bibinfo {author} {\bibfnamefont {A.}~\bibnamefont {McCray}}, \bibinfo
		{author} {\bibfnamefont {M.}~\bibnamefont {Randeria}}, \bibinfo {author}
		{\bibfnamefont {M.}~\bibnamefont {Naamneh}}, \bibinfo {author} {\bibfnamefont
			{A.}~\bibnamefont {Kanigel}},\ and\ \bibinfo {author} {\bibfnamefont {T.~R.}\
			\bibnamefont {Lemberger}},\ }\bibfield  {title} {\bibinfo {title} {{Evidence
				of two-dimensional quantum critical behavior in the superfluid density of
				extremely underdoped Bi${}_{2}$Sr${}_{2}$CaCu${}_{2}$O${}_{8+x}$}},\ }\href
	{https://doi.org/10.1103/PhysRevB.85.180507} {\bibfield  {journal} {\bibinfo
			{journal} {Phys. Rev. B}\ }\textbf {\bibinfo {volume} {85}},\ \bibinfo
		{pages} {180507(R)} (\bibinfo {year} {2012})}\BibitemShut {NoStop}%
	\bibitem [{\citenamefont {Zuev}\ \emph {et~al.}(2005)\citenamefont {Zuev},
		\citenamefont {Seog~Kim},\ and\ \citenamefont {Lemberger}}]{Zuev2005}%
	\BibitemOpen
	\bibfield  {author} {\bibinfo {author} {\bibfnamefont {Y.}~\bibnamefont
			{Zuev}}, \bibinfo {author} {\bibfnamefont {M.~S.}~\bibnamefont {Kim}},\
		and\ \bibinfo {author} {\bibfnamefont {T.~R.}\ \bibnamefont {Lemberger}},\
	}\bibfield  {title} {\bibinfo {title} {{Correlation between Superfluid
				Density and ${T}_{c}$ of Underdoped
				${\mathrm{YBa}}_{2}{\mathrm{Cu}}_{3}{\mathrm{O}}_{6+x}$ Near the
				Superconductor-Insulator Transition}},\ }\href
	{https://doi.org/10.1103/PhysRevLett.95.137002} {\bibfield  {journal}
		{\bibinfo  {journal} {Phys. Rev. Lett.}\ }\textbf {\bibinfo {volume} {95}},\
		\bibinfo {pages} {137002} (\bibinfo {year} {2005})}\BibitemShut {NoStop}%
	%
	%
	\bibitem [{\citenamefont {Medvedyeva}\ \emph {et~al.}(2000)\citenamefont
		{Medvedyeva}, \citenamefont {Kim},\ and\ \citenamefont
		{Minnhagen}}]{Medveyeva2000}%
	\BibitemOpen
	\bibfield  {author} {\bibinfo {author} {\bibfnamefont {K.}~\bibnamefont
			{Medvedyeva}}, \bibinfo {author} {\bibfnamefont {B.~J.}\ \bibnamefont
			{Kim}},\ and\ \bibinfo {author} {\bibfnamefont {P.}~\bibnamefont
			{Minnhagen}},\ }\bibfield  {title} {\bibinfo {title} {{Analysis of
				current-voltage characteristics of two-dimensional superconductors:
				Finite-size scaling behavior in the vicinity of the Kosterlitz-Thouless
				transition}},\ }\href {https://doi.org/10.1103/PhysRevB.62.14531} {\bibfield
		{journal} {\bibinfo  {journal} {Phys. Rev. B}\ }\textbf {\bibinfo {volume}
			{62}},\ \bibinfo {pages} {14531} (\bibinfo {year} {2000})}\BibitemShut
	{NoStop}%
	\bibitem [{\citenamefont {Ganguly}\ \emph {et~al.}(2015)\citenamefont
		{Ganguly}, \citenamefont {Chaudhuri}, \citenamefont {Raychaudhuri},\ and\
		\citenamefont {Benfatto}}]{Ganguly2015}%
	\BibitemOpen
	\bibfield  {author} {\bibinfo {author} {\bibfnamefont {R.}~\bibnamefont
			{Ganguly}}, \bibinfo {author} {\bibfnamefont {D.}~\bibnamefont {Chaudhuri}},
		\bibinfo {author} {\bibfnamefont {P.}~\bibnamefont {Raychaudhuri}},\ and\
		\bibinfo {author} {\bibfnamefont {L.}~\bibnamefont {Benfatto}},\ }\bibfield
	{title} {\bibinfo {title} {{Slowing down of vortex motion at the
				Berezinskii-Kosterlitz-Thouless transition in ultrathin NbN films}},\ }\href
	{https://doi.org/10.1103/PhysRevB.91.054514} {\bibfield  {journal} {\bibinfo
			{journal} {Phys. Rev. B}\ }\textbf {\bibinfo {volume} {91}},\ \bibinfo
		{pages} {054514} (\bibinfo {year} {2015})}\BibitemShut {NoStop}%
	\bibitem [{\citenamefont {Mallik}\ \emph {et~al.}(2022)\citenamefont {Mallik},
		\citenamefont {Ménard}, \citenamefont {Saïz}, \citenamefont {Witt},
		\citenamefont {Lesueur}, \citenamefont {Gloter}, \citenamefont {Benfatto},
		\citenamefont {Bibes},\ and\ \citenamefont {Bergeal}}]{Mallik2022}%
	\BibitemOpen
	\bibfield  {author} {\bibinfo {author} {\bibfnamefont {S.}~\bibnamefont
			{Mallik}}, \bibinfo {author} {\bibfnamefont {G.}~\bibnamefont {Ménard}},
		\bibinfo {author} {\bibfnamefont {G.}~\bibnamefont {Saïz}}, \bibinfo
		{author} {\bibfnamefont {H.}~\bibnamefont {Witt}}, \bibinfo {author}
		{\bibfnamefont {J.}~\bibnamefont {Lesueur}}, \bibinfo {author} {\bibfnamefont
			{A.}~\bibnamefont {Gloter}}, \bibinfo {author} {\bibfnamefont
			{L.}~\bibnamefont {Benfatto}}, \bibinfo {author} {\bibfnamefont
			{M.}~\bibnamefont {Bibes}},\ and\ \bibinfo {author} {\bibfnamefont
			{N.}~\bibnamefont {Bergeal}},\ }\bibfield  {title} {\bibinfo {title}
		{{Superfluid stiffness of a KTaO$_3$-based two-dimensional electron gas}},\
	}\bibfield  {journal} {\bibinfo  {journal} {arXiv:2204.09094}}
	(\bibinfo {year} {2022})\BibitemShut {NoStop}%
	\bibitem [{\citenamefont {Tamir}\ \emph {et~al.}(2019)\citenamefont {Tamir},
		\citenamefont {Benyamini}, \citenamefont {Telford}, \citenamefont
		{Gorniaczyk}, \citenamefont {Doron}, \citenamefont {Levinson}, \citenamefont
		{Wang}, \citenamefont {Gay}, \citenamefont {Sacepe}, \citenamefont {Hone},
		\citenamefont {Watanabe}, \citenamefont {Taniguchi}, \citenamefont {Dean},
		\citenamefont {Pasupathy},\ and\ \citenamefont {Shahar}}]{Tamir2019}%
	\BibitemOpen
	\bibfield  {author} {\bibinfo {author} {\bibfnamefont {I.}~\bibnamefont
			{Tamir}}, \bibinfo {author} {\bibfnamefont {A.}~\bibnamefont {Benyamini}},
		\bibinfo {author} {\bibfnamefont {E.~J.}\ \bibnamefont {Telford}}, \bibinfo
		{author} {\bibfnamefont {F.}~\bibnamefont {Gorniaczyk}}, \bibinfo {author}
		{\bibfnamefont {A.}~\bibnamefont {Doron}}, \bibinfo {author} {\bibfnamefont
			{T.}~\bibnamefont {Levinson}}, \bibinfo {author} {\bibfnamefont
			{D.}~\bibnamefont {Wang}}, \bibinfo {author} {\bibfnamefont {F.}~\bibnamefont
			{Gay}}, \bibinfo {author} {\bibfnamefont {B.}~\bibnamefont {Sacepe}},
		\bibinfo {author} {\bibfnamefont {J.}~\bibnamefont {Hone}}, \bibinfo {author}
		{\bibfnamefont {K.}~\bibnamefont {Watanabe}}, \bibinfo {author}
		{\bibfnamefont {T.}~\bibnamefont {Taniguchi}}, \bibinfo {author}
		{\bibfnamefont {C.~R.}\ \bibnamefont {Dean}}, \bibinfo {author}
		{\bibfnamefont {A.~N.}\ \bibnamefont {Pasupathy}},\ and\ \bibinfo {author}
		{\bibfnamefont {D.}~\bibnamefont {Shahar}},\ }\bibfield  {title} {\bibinfo
		{title} {Sensitivity of the superconducting state in thin films},\ }\bibfield
	{journal} {\bibinfo  {journal} {{Sci.~Advances}}\ }\textbf {\bibinfo
		{volume} {5}},\ (\bibinfo {year} {2019})\BibitemShut {NoStop}%
	\bibitem [{\citenamefont {Benyamini}\ \emph {et~al.}(2019)\citenamefont
		{Benyamini}, \citenamefont {Telford}, \citenamefont {Kennes}, \citenamefont
		{Wang}, \citenamefont {Williams}, \citenamefont {Watanabe}, \citenamefont
		{Taniguchi}, \citenamefont {Shahar}, \citenamefont {Hone}, \citenamefont
		{Dean}, \citenamefont {Millis},\ and\ \citenamefont
		{Pasupathy}}]{Benyamini2020}%
	\BibitemOpen
	\bibfield  {author} {\bibinfo {author} {\bibfnamefont {A.}~\bibnamefont
			{Benyamini}}, \bibinfo {author} {\bibfnamefont {E.~J.}\ \bibnamefont
			{Telford}}, \bibinfo {author} {\bibfnamefont {D.~M.}\ \bibnamefont {Kennes}},
		\bibinfo {author} {\bibfnamefont {D.}~\bibnamefont {Wang}}, \bibinfo {author}
		{\bibfnamefont {A.}~\bibnamefont {Williams}}, \bibinfo {author}
		{\bibfnamefont {K.}~\bibnamefont {Watanabe}}, \bibinfo {author}
		{\bibfnamefont {T.}~\bibnamefont {Taniguchi}}, \bibinfo {author}
		{\bibfnamefont {D.}~\bibnamefont {Shahar}}, \bibinfo {author} {\bibfnamefont
			{J.}~\bibnamefont {Hone}}, \bibinfo {author} {\bibfnamefont {C.~R.}\
			\bibnamefont {Dean}}, \bibinfo {author} {\bibfnamefont {A.~J.}\ \bibnamefont
			{Millis}},\ and\ \bibinfo {author} {\bibfnamefont {A.~N.}\ \bibnamefont
			{Pasupathy}},\ }\bibfield  {title} {\bibinfo {title} {Fragility of the
			dissipationless state in clean two-dimensional superconductors},\ }\href
	{https://doi.org/10.1038/s41567-019-0571-z} {\bibfield  {journal} {\bibinfo
			{journal} {{Nat.~Phys.}}\ }\textbf {\bibinfo {volume} {15}},\ \bibinfo
		{pages} {947} (\bibinfo {year} {2019})}\BibitemShut {NoStop}%
	\bibitem [{\citenamefont {Levinson}\ \emph {et~al.}(2019)\citenamefont
		{Levinson}, \citenamefont {Doron}, \citenamefont {Gorniaczyk},\ and\
		\citenamefont {Shahar}}]{Levinson2019}%
	\BibitemOpen
	\bibfield  {author} {\bibinfo {author} {\bibfnamefont {T.}~\bibnamefont
			{Levinson}}, \bibinfo {author} {\bibfnamefont {A.}~\bibnamefont {Doron}},
		\bibinfo {author} {\bibfnamefont {F.}~\bibnamefont {Gorniaczyk}},\ and\
		\bibinfo {author} {\bibfnamefont {D.}~\bibnamefont {Shahar}},\ }\bibfield
	{title} {\bibinfo {title} {Electron-phonon coupling across the
			superconductor-insulator transition},\ }\href
	{https://doi.org/10.1103/PhysRevB.100.184508} {\bibfield  {journal} {\bibinfo
			{journal} {{Phys.~Rev.~B}}\ }\textbf {\bibinfo {volume} {100}},\ \bibinfo
		{pages} {184508} (\bibinfo {year} {2019})}\BibitemShut {NoStop}%
	\bibitem [{\citenamefont {Linzen}\ \emph {et~al.}(2017)\citenamefont {Linzen},
		\citenamefont {Ziegler}, \citenamefont {Astafiev}, \citenamefont {Schmelz},
		\citenamefont {Hübner}, \citenamefont {Diegel}, \citenamefont {Il’ichev},\
		and\ \citenamefont {Meyer}}]{Linzen_2017}%
	\BibitemOpen
	\bibfield  {author} {\bibinfo {author} {\bibfnamefont {S.}~\bibnamefont
			{Linzen}}, \bibinfo {author} {\bibfnamefont {M.}~\bibnamefont {Ziegler}},
		\bibinfo {author} {\bibfnamefont {O.~V.}\ \bibnamefont {Astafiev}}, \bibinfo
		{author} {\bibfnamefont {M.}~\bibnamefont {Schmelz}}, \bibinfo {author}
		{\bibfnamefont {U.}~\bibnamefont {Hübner}}, \bibinfo {author} {\bibfnamefont
			{M.}~\bibnamefont {Diegel}}, \bibinfo {author} {\bibfnamefont
			{E.}~\bibnamefont {Il’ichev}},\ and\ \bibinfo {author} {\bibfnamefont
			{H.-G.}\ \bibnamefont {Meyer}},\ }\bibfield  {title} {\bibinfo {title}
		{Structural and electrical properties of ultrathin niobium nitride films
			grown by atomic layer deposition},\ }\href
	{https://doi.org/10.1088/1361-6668/aa572a} {\bibfield  {journal} {\bibinfo
			{journal} {Superconductor Science and Technology}\ }\textbf {\bibinfo
			{volume} {30}},\ \bibinfo {pages} {035010} (\bibinfo {year}
		{2017})}\BibitemShut {NoStop}%
	\bibitem [{\citenamefont {Baumgartner}\ \emph {et~al.}(2021)\citenamefont
		{Baumgartner}, \citenamefont {Fuchs}, \citenamefont {Fr\'esz}, \citenamefont
		{Reinhardt}, \citenamefont {Gronin}, \citenamefont {Gardner}, \citenamefont
		{Manfra}, \citenamefont {Paradiso},\ and\ \citenamefont
		{Strunk}}]{Baumgartner_2020}%
	\BibitemOpen
	\bibfield  {author} {\bibinfo {author} {\bibfnamefont {C.}~\bibnamefont
			{Baumgartner}}, \bibinfo {author} {\bibfnamefont {L.}~\bibnamefont {Fuchs}},
		\bibinfo {author} {\bibfnamefont {L.}~\bibnamefont {Fr\'esz}}, \bibinfo
		{author} {\bibfnamefont {S.}~\bibnamefont {Reinhardt}}, \bibinfo {author}
		{\bibfnamefont {S.}~\bibnamefont {Gronin}}, \bibinfo {author} {\bibfnamefont
			{G.~C.}\ \bibnamefont {Gardner}}, \bibinfo {author} {\bibfnamefont {M.~J.}\
			\bibnamefont {Manfra}}, \bibinfo {author} {\bibfnamefont {N.}~\bibnamefont
			{Paradiso}},\ and\ \bibinfo {author} {\bibfnamefont {C.}~\bibnamefont
			{Strunk}},\ }\bibfield  {title} {\bibinfo {title} {{Josephson inductance as a
				probe for highly ballistic semiconductor-superconductor weak links}},\ }\href
	{https://doi.org/10.1103/PhysRevLett.126.037001} {\bibfield  {journal}
		{\bibinfo  {journal} {Phys. Rev. Lett.}\ }\textbf {\bibinfo {volume} {126}},\
		\bibinfo {pages} {037001} (\bibinfo {year} {2021})}\BibitemShut {NoStop}%
	%
	\bibitem [{Sup()}]{Supplement}%
	\BibitemOpen
	\href@noop {} {\bibinfo {title} {{See Supplementary Information for further
				details.}}}\BibitemShut {Stop}%
	%
	\bibitem [{\citenamefont {Mondal}\ \emph
		{et~al.}(2011{\natexlab{b}})\citenamefont {Mondal}, \citenamefont
		{Kamlapure}, \citenamefont {Chand}, \citenamefont {Saraswat}, \citenamefont
		{Kumar}, \citenamefont {Jesudasan}, \citenamefont {Benfatto}, \citenamefont
		{Tripathi},\ and\ \citenamefont {Raychaudhuri}}]{Mondal_2011a}%
	\BibitemOpen
	\bibfield  {author} {\bibinfo {author} {\bibfnamefont {M.}~\bibnamefont
			{Mondal}}, \bibinfo {author} {\bibfnamefont {A.}~\bibnamefont {Kamlapure}},
		\bibinfo {author} {\bibfnamefont {M.}~\bibnamefont {Chand}}, \bibinfo
		{author} {\bibfnamefont {G.}~\bibnamefont {Saraswat}}, \bibinfo {author}
		{\bibfnamefont {S.}~\bibnamefont {Kumar}}, \bibinfo {author} {\bibfnamefont
			{J.}~\bibnamefont {Jesudasan}}, \bibinfo {author} {\bibfnamefont
			{L.}~\bibnamefont {Benfatto}}, \bibinfo {author} {\bibfnamefont
			{V.}~\bibnamefont {Tripathi}},\ and\ \bibinfo {author} {\bibfnamefont
			{P.}~\bibnamefont {Raychaudhuri}},\ }\bibfield  {title} {\bibinfo {title}
		{{Phase Fluctuations in a Strongly Disordered $s$-Wave {NbN} Superconductor
				Close to the Metal-Insulator Transition}},\ }\href
	{https://doi.org/10.1103/PhysRevLett.106.047001} {\bibfield  {journal}
		{\bibinfo  {journal} {Phys. Rev. Lett.}\ }\textbf {\bibinfo {volume} {106}},\
		\bibinfo {pages} {047001} (\bibinfo {year} {2011}{\natexlab{b}})}\BibitemShut
	{NoStop}%
	%
	\bibitem [{\citenamefont {Baturina}\ \emph {et~al.}(2012)\citenamefont
		{Baturina}, \citenamefont {Postolova}, \citenamefont {Mironov}, \citenamefont
		{Glatz}, \citenamefont {Baklanov},\ and\ \citenamefont
		{Vinokur}}]{Baturina_2012}%
	\BibitemOpen
	\bibfield  {author} {\bibinfo {author} {\bibfnamefont {T.~I.}\ \bibnamefont
			{Baturina}}, \bibinfo {author} {\bibfnamefont {S.~V.}\ \bibnamefont
			{Postolova}}, \bibinfo {author} {\bibfnamefont {A.~Y.}\ \bibnamefont
			{Mironov}}, \bibinfo {author} {\bibfnamefont {A.}~\bibnamefont {Glatz}},
		\bibinfo {author} {\bibfnamefont {M.~R.}\ \bibnamefont {Baklanov}},\ and\
		\bibinfo {author} {\bibfnamefont {V.~M.}\ \bibnamefont {Vinokur}},\
	}\bibfield  {title} {\bibinfo {title} {{Superconducting phase transitions in
				ultrathin {TiN} films}},\ }\href {https://doi.org/10.1209/0295-5075/97/17012}
	{\bibfield  {journal} {\bibinfo  {journal} {Europhys.~Lett.}\ }\textbf
		{\bibinfo {volume} {97}},\ \bibinfo {pages} {17012} (\bibinfo {year}
		{2012})}\BibitemShut {NoStop}%
	%
	%
	\bibitem [{\citenamefont {Postolova}\ \emph {et~al.}(2015)\citenamefont
		{Postolova}, \citenamefont {Mironov},\ and\ \citenamefont
		{Baturina}}]{Postolova2015}%
	\BibitemOpen
	\bibfield  {author} {\bibinfo {author} {\bibfnamefont {S.~V.}\ \bibnamefont
			{Postolova}}, \bibinfo {author} {\bibfnamefont {A.~Y.}\ \bibnamefont
			{Mironov}},\ and\ \bibinfo {author} {\bibfnamefont {T.~I.}\ \bibnamefont
			{Baturina}},\ }\bibfield  {title} {\bibinfo {title} {{Nonequilibrium transport
				near the superconducting transition in TiN films}},\ }\href
	{https://doi.org/10.1134/S0021364014220135} {\bibfield  {journal} {\bibinfo
			{journal} {JETP Letters}\ }\textbf {\bibinfo {volume} {100}},\ \bibinfo
		{pages} {635} (\bibinfo {year} {2015})}\BibitemShut {NoStop}%
	%
	\bibitem{mironov2018}
	A.Yu.~Mironov, S.V.~Postolova,  T.I.~Baturina, Quan\-tum contributions to the magnetoconductivity of cri\-tically disordered superconducting TiN films,
	J.~Physics: Cond.~Mat.~{\bf 40}, 485601 (2018)
%
\bibitem{Kronfeldner2021}
K.~Kronfeldner, T.~I.~Baturina, C.~Strunk, 
Multiple crossing points and possible quantum criticality in the magnetoresistance of thin TiN films,
Phys.~Rev.~B {\bf 103}, 184512 (2021).
	%
	\bibitem [{\citenamefont {Larkin}\ and\ \citenamefont
		{Varlamov}(2005)}]{LarkinVarlamov2005}%
	\BibitemOpen
	\bibfield  {author} {\bibinfo {author} {\bibfnamefont {A.~I.}\ \bibnamefont
			{Larkin}}\ and\ \bibinfo {author} {\bibfnamefont {A.}~\bibnamefont
			{Varlamov}},\ }\href@noop {} {\emph {\bibinfo {title} {{Theory of
					Fluctuations in Superconductors}}}}\ (\bibinfo  {publisher} {Clarendon Press,
		Oxford},\ \bibinfo {year} {2005})\BibitemShut {NoStop}%
	%
	\bibitem [{\citenamefont {Semenov}\ \emph {et~al.}(2009)\citenamefont
		{Semenov}, \citenamefont {G\"unther}, \citenamefont {B\"ottger},
		\citenamefont {H\"ubers}, \citenamefont {Bartolf}, \citenamefont {Engel},
		\citenamefont {Schilling}, \citenamefont {Ilin}, \citenamefont {Siegel},
		\citenamefont {Schneider}, \citenamefont {Gerthsen},\ and\ \citenamefont
		{Gippius}}]{Semenov2009}%
	\BibitemOpen
	\bibfield  {author} {\bibinfo {author} {\bibfnamefont {A.}~\bibnamefont
			{Semenov}}, \bibinfo {author} {\bibfnamefont {B.}~\bibnamefont {G\"unther}},
		\bibinfo {author} {\bibfnamefont {U.}~\bibnamefont {B\"ottger}}, \bibinfo
		{author} {\bibfnamefont {H.-W.}\ \bibnamefont {H\"ubers}}, \bibinfo {author}
		{\bibfnamefont {H.}~\bibnamefont {Bartolf}}, \bibinfo {author} {\bibfnamefont
			{A.}~\bibnamefont {Engel}}, \bibinfo {author} {\bibfnamefont
			{A.}~\bibnamefont {Schilling}}, \bibinfo {author} {\bibfnamefont
			{K.}~\bibnamefont {Ilin}}, \bibinfo {author} {\bibfnamefont {M.}~\bibnamefont
			{Siegel}}, \bibinfo {author} {\bibfnamefont {R.}~\bibnamefont {Schneider}},
		\bibinfo {author} {\bibfnamefont {D.}~\bibnamefont {Gerthsen}},\ and\
		\bibinfo {author} {\bibfnamefont {N.~A.}\ \bibnamefont {Gippius}},\
	}\bibfield  {title} {\bibinfo {title} {{Optical and transport properties of
				ultrathin NbN films and nanostructures}},\ }\href
	{https://doi.org/10.1103/PhysRevB.80.054510} {\bibfield  {journal} {\bibinfo
			{journal} {Phys. Rev. B}\ }\textbf {\bibinfo {volume} {80}},\ \bibinfo
		{pages} {054510} (\bibinfo {year} {2009})}\BibitemShut {NoStop}%
	%
	\bibitem{maccari2020}
	I.~Maccari, N.~Defenu,  L.~Benfatto, C.~Castellani, ~T.~Enss,
	Interplay of spin waves and vortices in the two-dimensional XY model at small vortex-core energy,
	Phys.~Rev.~B {\bf 102}, 104505 (2020).
	%
	\bibitem [{\citenamefont {K\"onig}\ \emph {et~al.}(2015)\citenamefont
		{K\"onig}, \citenamefont {Levchenko}, \citenamefont {Protopopov},
		\citenamefont {Gornyi}, \citenamefont {Burmistrov},\ and\ \citenamefont
		{Mirlin}}]{Koenig2015}%
	\BibitemOpen
	\bibfield  {author} {\bibinfo {author} {\bibfnamefont {E.~J.}\ \bibnamefont
			{K\"onig}}, \bibinfo {author} {\bibfnamefont {A.}~\bibnamefont {Levchenko}},
		\bibinfo {author} {\bibfnamefont {I.~V.}\ \bibnamefont {Protopopov}},
		\bibinfo {author} {\bibfnamefont {I.~V.}\ \bibnamefont {Gornyi}}, \bibinfo
		{author} {\bibfnamefont {I.~S.}\ \bibnamefont {Burmistrov}},\ and\ \bibinfo
		{author} {\bibfnamefont {A.~D.}\ \bibnamefont {Mirlin}},\ }\bibfield  {title}
	{\bibinfo {title} {Berezinskii-kosterlitz-thouless transition in
			homogeneously disordered superconducting films},\ }\href
	{https://doi.org/10.1103/PhysRevB.92.214503} {\bibfield  {journal} {\bibinfo
			{journal} {Phys. Rev. B}\ }\textbf {\bibinfo {volume} {92}},\ \bibinfo
		{pages} {214503} (\bibinfo {year} {2015})}\BibitemShut {NoStop}%
	\bibitem [{\citenamefont {Garland}\ and\ \citenamefont
		{Lee}(1987)}]{Garland1987}%
	\BibitemOpen
	\bibfield  {author} {\bibinfo {author} {\bibfnamefont {J.~C.}\ \bibnamefont
			{Garland}}\ and\ \bibinfo {author} {\bibfnamefont {H.~J.}\ \bibnamefont
			{Lee}},\ }\bibfield  {title} {\bibinfo {title} {Influence of a magnetic field
			on the two-dimensional phase transition in thin-film superconductors},\
	}\href {https://doi.org/10.1103/PhysRevB.36.3638} {\bibfield  {journal}
		{\bibinfo  {journal} {Phys. Rev. B}\ }\textbf {\bibinfo {volume} {36}},\
		\bibinfo {pages} {3638} (\bibinfo {year} {1987})}\BibitemShut {NoStop}%
	\bibitem [{\citenamefont {Minnhagen}(1984)}]{Minnhagen1984}%
	\BibitemOpen
	\bibfield  {author} {\bibinfo {author} {\bibfnamefont {P.}~\bibnamefont
			{Minnhagen}},\ }\bibfield  {title} {\bibinfo {title} {Evidence of magnetic
			field scaling for two-dimensional superconductors},\ }\href
	{https://doi.org/10.1103/PhysRevB.29.1440} {\bibfield  {journal} {\bibinfo
			{journal} {Phys. Rev. B}\ }\textbf {\bibinfo {volume} {29}},\ \bibinfo
		{pages} {1440} (\bibinfo {year} {1984})}\BibitemShut {NoStop}%
	\bibitem [{\citenamefont {Benfatto}\ \emph {et~al.}(2013)\citenamefont
		{Benfatto}, \citenamefont {Castellani},\ and\ \citenamefont
		{Giamarchi}}]{Benfatto_2013}%
	\BibitemOpen
	\bibfield  {author} {\bibinfo {author} {\bibfnamefont {L.}~\bibnamefont
			{Benfatto}}, \bibinfo {author} {\bibfnamefont {C.}~\bibnamefont
			{Castellani}},\ and\ \bibinfo {author} {\bibfnamefont {T.}~\bibnamefont
			{Giamarchi}},\ }\bibfield  {title} {\bibinfo {title}
		{{B}erezinskii–{K}osterlitz–{T}houless transition within the
			sine-{G}ordon approach: The role of the vortex-core energy},\ }\href
	{https://doi.org/10.1142/9789814417648_0005} {\bibfield  {journal} {\bibinfo
			{journal} {40 Years of Berezinskii–Kosterlitz–Thouless Theory}\ ,\
			\bibinfo {pages} {161–199}} (\bibinfo {year} {2013})}
	\BibitemShut {NoStop}%
	%
	\bibitem [{\citenamefont {Maccari}\ \emph {et~al.}(2017)\citenamefont
		{Maccari}, \citenamefont {Benfatto},\ and\ \citenamefont
		{Castellani}}]{Maccari2017}%
	\BibitemOpen
	\bibfield  {author} {\bibinfo {author} {\bibfnamefont {I.}~\bibnamefont
			{Maccari}}, \bibinfo {author} {\bibfnamefont {L.}~\bibnamefont {Benfatto}},\
		and\ \bibinfo {author} {\bibfnamefont {C.}~\bibnamefont {Castellani}},\
	}\bibfield  {title} {\bibinfo {title} {{Broadening of the
				Berezinskii-Kosterlitz-Thouless transition by correlated disorder}},\ }\href
	{https://doi.org/10.1103/PhysRevB.96.060508} {\bibfield  {journal} {\bibinfo
			{journal} {Phys. Rev. B}\ }\textbf {\bibinfo {volume} {96}},\ \bibinfo
		{pages} {060508(R)} (\bibinfo {year} {2017})}\BibitemShut {NoStop}%
	%
	\bibitem [{\citenamefont {Kamlapure}\ \emph {et~al.}(2013)\citenamefont
		{Kamlapure}, \citenamefont {Das}, \citenamefont {Ganguli}, \citenamefont
		{Parmar}, \citenamefont {Bhattacharyya},\ and\ \citenamefont
		{Raychaudhuri}}]{Kamlapure2013}%
	\BibitemOpen
	\bibfield  {author} {\bibinfo {author} {\bibfnamefont {A.}~\bibnamefont
			{Kamlapure}}, \bibinfo {author} {\bibfnamefont {T.}~\bibnamefont {Das}},
		\bibinfo {author} {\bibfnamefont {S.~C.}\ \bibnamefont {Ganguli}}, \bibinfo
		{author} {\bibfnamefont {J.~B.}\ \bibnamefont {Parmar}}, \bibinfo {author}
		{\bibfnamefont {S.}~\bibnamefont {Bhattacharyya}},\ and\ \bibinfo {author}
		{\bibfnamefont {P.}~\bibnamefont {Raychaudhuri}},\ }\bibfield  {title}
	{\bibinfo {title} {Emergence of nanoscale inhomogeneity in the
			superconducting state of a homogeneously disordered conventional
			superconductor},\ }\href {https://doi.org/10.1038/srep02979} {\bibfield
		{journal} {\bibinfo  {journal} {Scientific Reports}\ }\textbf {\bibinfo
			{volume} {3}},\ \bibinfo {pages} {2979} (\bibinfo {year} {2013})}\BibitemShut
	{NoStop}%
	%
	\bibitem{Maccari2018}
	I.~Maccari, L.~Benfatto, C.~Castellani, 
	The BKT Universality Class in the Presence of Correlated Disorder,
	Condens.~Matt.~{\bf 3}, 8 (2018).
	%
	\bibitem{Maccari2019}
	I.~Maccari, L.~Benfatto, C.~Castellani, 
	Disordered XY model: Effective medium theory and beyond,
	Phys.~Rev.~B {\bf 99}, 104509 (2019).  
	%
	\bibitem [{\citenamefont {Reyren}\ \emph {et~al.}(2007)\citenamefont {Reyren},
		\citenamefont {Thiel}, \citenamefont {Caviglia}, \citenamefont {Kourkoutis},
		\citenamefont {Hammerl}, \citenamefont {Richter}, \citenamefont {Schneider},
		\citenamefont {Kopp}, \citenamefont {Rüetschi}, \citenamefont {Jaccard},
		\citenamefont {Gabay}, \citenamefont {Muller}, \citenamefont {Triscone},\
		and\ \citenamefont {Mannhart}}]{Reyren2009}%
	\BibitemOpen
	\bibfield  {author} {\bibinfo {author} {\bibfnamefont {N.}~\bibnamefont
			{Reyren}}, \bibinfo {author} {\bibfnamefont {S.}~\bibnamefont {Thiel}},
		\bibinfo {author} {\bibfnamefont {A.~D.}\ \bibnamefont {Caviglia}}, \bibinfo
		{author} {\bibfnamefont {L.~F.}\ \bibnamefont {Kourkoutis}}, \bibinfo
		{author} {\bibfnamefont {G.}~\bibnamefont {Hammerl}}, \bibinfo {author}
		{\bibfnamefont {C.}~\bibnamefont {Richter}}, \bibinfo {author} {\bibfnamefont
			{C.~W.}\ \bibnamefont {Schneider}}, \bibinfo {author} {\bibfnamefont
			{T.}~\bibnamefont {Kopp}}, \bibinfo {author} {\bibfnamefont {A.-S.}\
			\bibnamefont {Rüetschi}}, \bibinfo {author} {\bibfnamefont {D.}~\bibnamefont
			{Jaccard}}, \bibinfo {author} {\bibfnamefont {M.}~\bibnamefont {Gabay}},
		\bibinfo {author} {\bibfnamefont {D.~A.}\ \bibnamefont {Muller}}, \bibinfo
		{author} {\bibfnamefont {J.-M.}\ \bibnamefont {Triscone}},\ and\ \bibinfo
		{author} {\bibfnamefont {J.}~\bibnamefont {Mannhart}},\ }\bibfield  {title}
	{\bibinfo {title} {Superconducting interfaces between insulating oxides},\
	} {\bibfield  {journal}
		{\bibinfo  {journal} {Science}\ }\textbf {\bibinfo {volume} {317}},\ \bibinfo
		{pages} {1196} (\bibinfo {year} {2007})}\  \BibitemShut
	{NoStop}%
	\bibitem [{\citenamefont {{\it et al.}}(2014)}]{Zhang_2014}%
	\BibitemOpen
	\bibfield  {author} {\bibinfo {author} {\bibfnamefont {Z.~W.-H.}\
			\bibnamefont {{\it et al.}}},\ }\bibfield  {title} {\bibinfo {title} {Direct
			observation of high-temperature superconductivity in one-unit-cell fese
			films},\ }\href {https://doi.org/10.1088/0256-307X/31/1/017401} {\bibfield
		{journal} {\bibinfo  {journal} {Chin.~Phys.~Lett.}\ }\textbf {\bibinfo
			{volume} {31}},\ \bibinfo {pages} {017401} (\bibinfo {year}
		{2014})}\BibitemShut {NoStop}%
	\bibitem [{\citenamefont {Lu}\ \emph {et~al.}(2015)\citenamefont {Lu},
		\citenamefont {Zheliuk}, \citenamefont {Leermakers}, \citenamefont {Yuan},
		\citenamefont {Zeitler}, \citenamefont {Law},\ and\ \citenamefont
		{Ye}}]{Lu_2015}%
	\BibitemOpen
	\bibfield  {author} {\bibinfo {author} {\bibfnamefont {J.~M.}\ \bibnamefont
			{Lu}}, \bibinfo {author} {\bibfnamefont {O.}~\bibnamefont {Zheliuk}},
		\bibinfo {author} {\bibfnamefont {I.}~\bibnamefont {Leermakers}}, \bibinfo
		{author} {\bibfnamefont {N.~F.~Q.}\ \bibnamefont {Yuan}}, \bibinfo {author}
		{\bibfnamefont {U.}~\bibnamefont {Zeitler}}, \bibinfo {author} {\bibfnamefont
			{K.~T.}\ \bibnamefont {Law}},\ and\ \bibinfo {author} {\bibfnamefont {J.~T.}\
			\bibnamefont {Ye}},\ }\bibfield  {title} {\bibinfo {title} {{Evidence for
				two-dimensional Ising superconductivity in gated MoS$_2$}},\ }
	{\bibfield  {journal} {\bibinfo
			{journal} {Science}\ }\textbf {\bibinfo {volume} {350}},\ \bibinfo {pages}
		{1353} (\bibinfo {year} {2015})},\  \BibitemShut
	{NoStop}%
	\bibitem [{\citenamefont {Cao}\ \emph {et~al.}(2018)\citenamefont {Cao},
		\citenamefont {Fatemi}, \citenamefont {Fang}, \citenamefont {Watanabe},
		\citenamefont {Taniguchi}, \citenamefont {Kaxiras},\ and\ \citenamefont
		{Jarillo-Herrero}}]{cao_super_2018}%
	\BibitemOpen
	\bibfield  {author} {\bibinfo {author} {\bibfnamefont {Y.}~\bibnamefont
			{Cao}}, \bibinfo {author} {\bibfnamefont {V.}~\bibnamefont {Fatemi}},
		\bibinfo {author} {\bibfnamefont {S.}~\bibnamefont {Fang}}, \bibinfo {author}
		{\bibfnamefont {K.}~\bibnamefont {Watanabe}}, \bibinfo {author}
		{\bibfnamefont {T.}~\bibnamefont {Taniguchi}}, \bibinfo {author}
		{\bibfnamefont {E.}~\bibnamefont {Kaxiras}},\ and\ \bibinfo {author}
		{\bibfnamefont {P.}~\bibnamefont {Jarillo-Herrero}},\ }\bibfield  {title}
	{\bibinfo {title} {Unconventional superconductivity in magic-angle graphene
			superlattices},\ }\href {https://doi.org/10.1038/nature26160} {\bibfield
		{journal} {\bibinfo  {journal} {Nature}\ }\textbf {\bibinfo {volume} {556}},\
		\bibinfo {pages} {43} (\bibinfo {year} {2018})}\BibitemShut {NoStop}%
	\bibitem [{\citenamefont {Park}\ \emph {et~al.}(2021)\citenamefont {Park},
		\citenamefont {Cao}, \citenamefont {Watanabe}, \citenamefont {Taniguchi},\
		and\ \citenamefont {Jarillo-Herrero}}]{Park_2021}%
	\BibitemOpen
	\bibfield  {author} {\bibinfo {author} {\bibfnamefont {J.~M.}\ \bibnamefont
			{Park}}, \bibinfo {author} {\bibfnamefont {Y.}~\bibnamefont {Cao}}, \bibinfo
		{author} {\bibfnamefont {K.}~\bibnamefont {Watanabe}}, \bibinfo {author}
		{\bibfnamefont {T.}~\bibnamefont {Taniguchi}},\ and\ \bibinfo {author}
		{\bibfnamefont {P.}~\bibnamefont {Jarillo-Herrero}},\ }\bibfield  {title}
	{\bibinfo {title} {Tunable strongly coupled superconductivity in magic-angle
			twisted trilayer graphene},\ }\href
	{https://doi.org/10.1038/s41586-021-03192-0} {\bibfield  {journal} {\bibinfo
			{journal} {Nature}\ }\textbf {\bibinfo {volume} {590}},\ \bibinfo {pages}
		{249} (\bibinfo {year} {2021})}\BibitemShut {NoStop}%
	\bibitem [{\citenamefont {Zhou}\ \emph {et~al.}(2021)\citenamefont {Zhou},
		\citenamefont {Xie}, \citenamefont {Taniguchi}, \citenamefont {Watanabe},\
		and\ \citenamefont {Young}}]{Zhou_Young_2021}%
	\BibitemOpen
	\bibfield  {author} {\bibinfo {author} {\bibfnamefont {H.}~\bibnamefont
			{Zhou}}, \bibinfo {author} {\bibfnamefont {T.}~\bibnamefont {Xie}}, \bibinfo
		{author} {\bibfnamefont {T.}~\bibnamefont {Taniguchi}}, \bibinfo {author}
		{\bibfnamefont {K.}~\bibnamefont {Watanabe}},\ and\ \bibinfo {author}
		{\bibfnamefont {A.~F.}\ \bibnamefont {Young}},\ }\bibfield  {title} {\bibinfo
		{title} {Superconductivity in rhombohedral trilayer graphene},\ }\href
	{https://doi.org/10.1038/s41586-021-03926-0} {\bibfield  {journal} {\bibinfo
			{journal} {Nature}\ }\textbf {\bibinfo {volume} {598}},\ \bibinfo {pages}
		{434} (\bibinfo {year} {2021})}\BibitemShut {NoStop}
	%
	\bibitem{Enze_Zhang2023}
	E.~Zhang, Y.~M.~Xie, Y.~Q.~Fang,~J.~L.~Zhang, X.~Xu, Y.~C.~Zou, P.~L.~Leng, ~X.~J.~Gao, Y.~Zhang, L.~F.~Ai, Y.~D.~Zhang, Z.~H.~Jia, S.~S.~Liu, J.~Y.~Yan, W.~Zhao, S.~Haigh, X.~F.~Kou, J.~S.~Yang, F.~Q.~Huang, K.~T.~Law, F.~X.~Xiu, and S.~M.~Dong, Spin–orbit–parity coupled superconductivity in atomically thin 2M-WS$_2
	$,  Nat.~Phys.~{\bf 19}, 106 (2023).
	%
	\bibitem [{\citenamefont {Lopes~dos Santos}\ and\ \citenamefont
		{Abrahams}(1985)}]{SantosAbrahams1985}%
	\BibitemOpen
	\bibfield  {author} {\bibinfo {author} {\bibfnamefont {J.~M.~B.}\
			\bibnamefont {Lopes~dos Santos}}\ and\ \bibinfo {author} {\bibfnamefont
			{E.}~\bibnamefont {Abrahams}},\ }\bibfield  {title} {\bibinfo {title}
		{Superconducting fluctuation conductivity in a magnetic field in two
			dimensions},\ }\href {https://doi.org/10.1103/PhysRevB.31.172} {\bibfield
		{journal} {\bibinfo  {journal} {Phys. Rev. B}\ }\textbf {\bibinfo {volume}
			{31}},\ \bibinfo {pages} {172} (\bibinfo {year} {1985})}\BibitemShut
	{NoStop}%
	\bibitem [{\citenamefont {Tinkham}(1996)}]{Tinkham1996}%
	\BibitemOpen
	\bibfield  {author} {\bibinfo {author} {\bibfnamefont {M.}~\bibnamefont
			{Tinkham}},\ }\href@noop {} {\emph {\bibinfo {title} {{Introduction to
					Superconductivity}}}}\ (\bibinfo  {publisher} {McGraw-Hill},\ \bibinfo {year}
	{1996})\BibitemShut {NoStop}%
\end{thebibliography}
\end{document}